\documentclass[nofootinbib,prd]{revtex4}%
\usepackage{amsmath}
\usepackage{amsfonts}
\usepackage{amssymb}
\usepackage{graphicx}%
\begin{document}
\title{Distorting General Relativity: Gravity's Rainbow and $f(R)$ theories at work}
\author{Remo Garattini}
\email{Remo.Garattini@unibg.it}
\affiliation{Universit\`{a} degli Studi di Bergamo, Facolt\`{a} di Ingegneria,}
\affiliation{Viale Marconi 5, 24044 Dalmine (Bergamo) Italy}
\affiliation{I.N.F.N. - sezione di Milano, Milan, Italy.}

\begin{abstract}
We compute the Zero Point Energy in a spherically symmetric background
combining the high energy distortion of Gravity's Rainbow with the
modification induced by a f(R) theory. Here f(R) is a generic analytic
function of the Ricci curvature scalar R in 4D and in 3D. The explicit
calculation is performed for a Schwarzschild metric. Due to the spherically
symmetric property of the Schwarzschild metric we can compare the effects of
the modification induced by a f(R) theory in 4D and in 3D. We find that the
final effect of the combined theory is to have finite quantities that shift
the Zero Point Energy. In this context we setup a Sturm-Liouville problem with
the cosmological constant considered as the associated eigenvalue. The
eigenvalue equation is a reformulation of the Wheeler-DeWitt equation which is
analyzed by means of a variational approach based on gaussian trial
functionals. With the help of a canonical decomposition, we find that the
relevant contribution to one loop is given by the graviton quantum
fluctuations around the given background. A final discussion on the connection
of our result with the observed cosmological constant is also reported.

\end{abstract}
\maketitle

\section{Introduction}

In recent years many efforts have been done to modify General Relativity under
different aspects. One of the main reasons is the search for a satisfying
model of inflation and from another point of view try to find a
\textit{Quantum Gravity theory} which is still lacking. From one side, some
modifications have their foundation in a class of theories termed
\textit{Extended Theories of Gravity} (ETG) which have become a sort of
paradigm in the study of gravitational interaction based on corrections and
enlargements of the Einstein scheme. The paradigm consists, essentially, in
adding higher-order curvature invariants and non-minimally coupled scalar
fields into dynamics resulting from the effective action of Quantum Gravity
\cite{BOS}. Such corrective terms seem to be unavoidable if we want to obtain
the effective action of Quantum Gravity on scales closed to the Planck length
\cite{vilkovisky}. Therefore terms of the form $\mathcal{R}^{2}$,
$\mathcal{R}^{\mu\nu}\mathcal{R}_{\mu\nu}$, $\mathcal{R}^{\mu\nu\alpha\beta
}\mathcal{R}_{\mu\nu\alpha\beta}$, $\mathcal{R}\,\Box\mathcal{R}$, or
$\mathcal{R}\,\Box^{k}\mathcal{R}$ have to be added to the effective
Lagrangian of gravitational field when quantum corrections are considered. To
this purpose one should consider the possibility that the Hilbert\thinspace
-\thinspace Einstein Lagrangian, linear in the scalar curvature $\mathcal{R}$,
should be generalized into a generic function $f(\mathcal{R})$\footnote{For a
recent review, see Refs.\cite{SOSO,SCMdL,TSVF}.}. Of course this modification
will have some effects not only on large scale structure of space time, but
even at small scales where quantum effects come into play. On the other side,
one can distort space time introducing a modification which activates when the
Planck scale energy is approached. Such a modification in its simplest form is
known as \textit{Doubly Special Relativity} or \textit{Deformed Special
Relativity} (DSR)\cite{GAC:2000,GAC:2000ge,DSR}\textbf{.} One of the
characterizing DSR effects is that the usual dispersion relation of a massive
particle of mass $m$ is modified into the following expression%
\begin{equation}
E^{2}g_{1}^{2}\left(  E/E_{P}\right)  -p^{2}g_{2}^{2}\left(  E/E_{P}\right)
=m^{2},\label{mdisp}%
\end{equation}
where $g_{1}\left(  E/E_{P}\right)  $ and $g_{2}\left(  E/E_{P}\right)  $ are
two arbitrary functions which have the following property%
\begin{equation}
\lim_{E/E_{P}\rightarrow0}g_{1}\left(  E/E_{P}\right)  =1\qquad\text{and}%
\qquad\lim_{E/E_{P}\rightarrow0}g_{2}\left(  E/E_{P}\right)  =1.\label{lim}%
\end{equation}
The low energy limit $\left(  \ref{lim}\right)  $ ensures that the usual
dispersion relation is recovered. An immediate generalization to a curved
background led Magueijo and Smolin\cite{MagSmo} to introduce the idea of
\textit{Gravity's Rainbow }where a one parameter family of equations%
\begin{equation}
G_{\mu\nu}\left(  E\right)  =8\pi G\left(  E\right)  T_{\mu\nu}\left(
E\right)  +g_{\mu\nu}\Lambda\left(  E\right)  ,\label{Gmn}%
\end{equation}
replaces the ordinary Einstein's Field Equations. The meaning of $G\left(
E\right)  $ is represented by an energy dependent Newton's constant, defined
so that $G\left(  0\right)  $ is the low-energy Newton's constant and
similarly we have an energy dependent cosmological constant $\Lambda\left(
E\right)  $, defined so that $\Lambda\left(  0\right)  $ is the low-energy
cosmological constant. Solutions of $\left(  \ref{Gmn}\right)  $ have been
found in Ref.\cite{MagSmo}. One of these is the distorted Schwarzschild
solution%
\begin{equation}
ds^{2}=-\left(  1-\frac{2MG\left(  0\right)  }{r}\right)  \frac{dt^{2}}%
{g_{1}^{2}\left(  E/E_{P}\right)  }+\frac{dr^{2}}{\left(  1-\frac{2MG\left(
0\right)  }{r}\right)  g_{2}^{2}\left(  E/E_{P}\right)  }+\frac{r^{2}}%
{g_{2}^{2}\left(  E/E_{P}\right)  }\left(  d\theta^{2}+\sin^{2}\theta
d\phi^{2}\right)  ,\label{line}%
\end{equation}
whose generalization leads to the following \textit{rainbow} version of a
spherically symmetric line element%
\begin{equation}
ds^{2}=-\frac{N^{2}\left(  r\right)  }{g_{1}^{2}\left(  E/E_{P}\right)
}dt^{2}+\frac{dr^{2}}{\left(  1-\frac{b\left(  r\right)  }{r}\right)
g_{2}^{2}\left(  E/E_{P}\right)  }+\frac{r^{2}}{g_{2}^{2}\left(
E/E_{P}\right)  }\left(  d\theta^{2}+\sin^{2}\theta d\phi^{2}\right)
.\label{dS}%
\end{equation}
$N$ is known as the lapse function and $b\left(  r\right)  $ is subject to the
only condition $b\left(  r_{t}\right)  =r_{t}=2MG$. Motivated by the previous
results obtained in $f(\mathcal{R})$ theories discussed in Ref.\cite{CG} and
in Gravity's Rainbow obtained in\footnote{Application of Gravity's Rainbow to
discuss the traversability of wormholes can be found in Ref.\cite{RGFL}.}
Refs.\cite{GaMa, GaMa1, RemoPLB}, in this paper we address the problem of
computing Zero Point Energy (ZPE) combining the two modifications. It is known
that ZPE calculations are affected by divergences that usually are kept under
control with the help of a regularization and renormalization procedure. From
this side, with an appropriate choice of the functions $g_{1}\left(
E/E_{P}\right)  $ and $g_{2}\left(  E/E_{P}\right)  $, Gravity's Rainbow seems
to offer a method which avoids the usual regularization and renormalization
procedure. Thus, while Gravity's Rainbow comes into play at Planckian scales,
$f(\mathcal{R})$ theories are related to the large scale structure of space
time, even if effects on short scales can be relevant, especially for ZPE
evaluations\cite{CG}. In ordinary gravity the computation of ZPE for quantum
fluctuations of the \textit{pure gravitational field} can be extracted by
rewriting the Wheeler-DeWitt equation (WDW)\cite{DeWitt} in a form which looks
like an expectation value computation\cite{Remo}. Its derivation is a
consequence of the Arnowitt-Deser-Misner ($\mathcal{ADM}$)
decomposition\cite{ADM} of space time based on the following line element%
\begin{equation}
ds^{2}=g_{\mu\nu}\left(  x\right)  dx^{\mu}dx^{\nu}=\left(  -N^{2}+N_{i}%
N^{i}\right)  dt^{2}+2N_{j}dtdx^{j}+g_{ij}dx^{i}dx^{j},
\end{equation}
where $N$ is the \textit{lapse }function and $N_{i}$ the \textit{shift
}function. In terms of the $\mathcal{ADM}$ variables, the four dimensional
scalar curvature $\mathcal{R}$ can be decomposed in the following way%
\begin{equation}
\mathcal{R}=R+K_{ij}K^{ij}-\left(  K\right)  ^{2}-2\nabla_{\mu}\left(
Ku^{\mu}+a^{\mu}\right)  ,\label{R}%
\end{equation}
where%
\begin{equation}
K_{ij}=-\frac{1}{2N}\left[  \partial_{t}g_{ij}-N_{i|j}-N_{j|i}\right]
\end{equation}
is the second fundamental form, $K=$ $g^{ij}K_{ij}$ is its trace, $R$ is the
three dimensional scalar curvature and $\sqrt{g}$ is the three dimensional
determinant of the metric. The last term in $\left(  \ref{R}\right)  $
represents the boundary terms contribution where the four-velocity $u^{\mu}$
is the timelike unit vector normal to the spacelike hypersurfaces (t=constant)
denoted by $\Sigma_{t}$ and $a^{\mu}=u^{\alpha}\nabla_{\alpha}u^{\mu}$ is the
acceleration of the timelike normal $u^{\mu}$. Thus%
\begin{equation}
\mathcal{L}\left[  N,N_{i},g_{ij}\right]  =\sqrt{-\text{\/\thinspace
\thinspace}^{4}\text{\/{}\negthinspace}g}\left(  \mathcal{R}-2\Lambda\right)
=\frac{N}{2\kappa}\sqrt{g}\text{ }\left[  K_{ij}K^{ij}-K^{2}+\,R-2\Lambda
-2\nabla_{\mu}\left(  Ku^{\mu}+a^{\mu}\right)  \right]  \label{Lag}%
\end{equation}
represents the gravitational Lagrangian density where $\kappa=8\pi G$ with $G$
the Newton's constant and for the sake of generality we have also included a
cosmological constant $\Lambda$. After a Legendre transformation, the WDW
equation simply becomes%
\begin{equation}
\mathcal{H}\Psi=\left[  \left(  2\kappa\right)  G_{ijkl}\pi^{ij}\pi^{kl}%
-\frac{\sqrt{g}}{2\kappa}\!{}\!\left(  \,\!R-2\Lambda\right)  \right]
\Psi=0,\label{WDWO}%
\end{equation}
where $G_{ijkl}$ is the super-metric and where the conjugate super-momentum
$\pi^{ij}$ is defined as%
\begin{equation}
\pi^{ij}=\frac{\delta\mathcal{L}}{\delta\left(  \partial_{t}g_{ij}\right)
}=\left(  g^{ij}K-K^{ij}\text{ }\right)  \frac{\sqrt{g}}{2\kappa}.\label{mom}%
\end{equation}
Note that $\mathcal{H}=0$ represents the classical constraint which guarantees
the invariance under time reparametrization. The other classical constraint
represents the invariance by spatial diffeomorphism and it is described by
$\pi_{|j}^{ij}=0$, where the vertical stroke \textquotedblleft%
$\vert$%
\textquotedblright\ denotes the covariant derivative with respect to the $3D$
metric $g_{ij}$. Formally, the WDW equation can be transformed into an
eigenvalue equation if we multiply Eq.$\left(  \ref{WDWO}\right)  $ by
$\Psi^{\ast}\left[  g_{ij}\right]  $ and functionally integrate over the three
spatial metric $g_{ij}$. What we obtain is\cite{Remo}\footnote{An application
of this calculation in the framework of Hor\v{a}va-Lifshits theory can be
found in Ref.\cite{RemoHL}}%
\begin{equation}
\frac{1}{V}\frac{\int\mathcal{D}\left[  g_{ij}\right]  \Psi^{\ast}\left[
g_{ij}\right]  \int_{\Sigma}d^{3}x\hat{\Lambda}_{\Sigma}\Psi\left[
g_{ij}\right]  }{\int\mathcal{D}\left[  g_{ij}\right]  \Psi^{\ast}\left[
g_{ij}\right]  \Psi\left[  g_{ij}\right]  }=\frac{1}{V}\frac{\left\langle
\Psi\left\vert \int_{\Sigma}d^{3}x\hat{\Lambda}_{\Sigma}\right\vert
\Psi\right\rangle }{\left\langle \Psi|\Psi\right\rangle }=-\frac{\Lambda
}{\kappa},\label{VEVO}%
\end{equation}
where we have also integrated over the hypersurface $\Sigma$ and we have
defined%
\begin{equation}
V=\int_{\Sigma}d^{3}x\sqrt{g}\label{Vol}%
\end{equation}
as the volume of the hypersurface $\Sigma$ with%
\begin{equation}
\hat{\Lambda}_{\Sigma}=\left(  2\kappa\right)  G_{ijkl}\pi^{ij}\pi^{kl}%
-\sqrt{g}R/\left(  2\kappa\right)  .\label{LambdaSigma}%
\end{equation}
In this form, Eq.$\left(  \ref{VEVO}\right)  $ can be used to compute ZPE
provided that $\Lambda/\kappa$ be considered as an eigenvalue of $\hat
{\Lambda}_{\Sigma}$, namely the WDW equation is transformed into an
expectation value computation. Nevertheless, solving Eq.$\left(
\ref{VEV}\right)  $ is a quite impossible task, therefore we are oriented to
use a variational approach with trial wave functionals. The related boundary
conditions are dictated by the choice of the trial wave functionals which, in
our case\textbf{,} are of the Gaussian type: this choice is justified by the
fact that ZPE should be described by a good candidate of the \textquotedblleft%
\textit{vacuum state}\textquotedblright. However if we change the form of the
wave functionals we change also the corresponding boundary conditions and
therefore the description of the vacuum state. It is better to observe that
the obtained eigenvalue $\Lambda/\kappa$, it is far to be a constant, rather
it will be dependent on some parameters like the mass $M$ and the radial
coordinate $r$ for the Schwarzschild case and therefore it will be considered
more like a \textquotedblleft\textit{dynamical cosmological constant}%
\textquotedblright\ evolving in $r$ and $M$ instead of a temporal parameter
$t$. This is not a novelty, almost all the inflationary models try to
substitute a cosmological constant $\Lambda$ with some fields that change with
time. In this case, it is the gravity itself that gives a dynamical aspect to
the \textquotedblleft\textit{cosmological constant} $\Lambda$%
\textquotedblright, or more correctly the ZPE $\Lambda/\kappa$, without
introducing any kind of external field but only quantum fluctuations. It
remains to say that the additional distortion due to a $f(\mathcal{R})$ metric
can be examined from two similar, but different point of view. Indeed for a
spherically symmetric background, thanks to the $\mathcal{ADM}$ decomposition
one finds the identity $\mathcal{R}=R$ as can be seen from Eq.$\left(
\ref{R}\right)  $. Thus in this paper we will consider the effects on a ZPE
computation of both a $f(\mathcal{R})$ theory as well as the ones of a $f(R)$
theory, namely a sub-class of the full covariant $f(\mathcal{R})$ models. It
is relevant to say that models of $f(R)$ theories has been considered recently
in the context of the Ho\v{r}ava-Lifshitz gravity\cite{HL}, where the symmetry
between space and time is explicitly broken. However both of the two
formulations joined to Gravity's Rainbow have to pass some tests. One of these
is the Minkowski limit: in other words when the gravitational field is
switched off, the induced cosmological constant must vanish. An example of
this behavior can be found in Yang-Mills theory where the energy density for
an $SU\left(  2\right)  $ massless gluon field in a constant color magnetic
field can be written\cite{NO}
\begin{equation}
\frac{E}{V}=\frac{1}{2}H^{2}+\frac{11}{48\pi^{2}}\left(  eH\right)  ^{2}%
\ln\left(  \frac{eH}{\mu^{2}}-\frac{1}{2}\right)  .
\end{equation}
This expression has a minimum away from the point $H=0$, namely
\begin{equation}
eH_{\min}=\mu^{2}\exp\left(  -\frac{24\pi^{2}}{11e^{2}}\right)
\end{equation}
and when $H\rightarrow0$, $E/V\rightarrow0$ that it means that in absence of
an external field, the energy density is absent. The transposition to the
gravitational field even if it is not immediate, it is pertinent: the
gravitational field is the analogue of the Yang-Mills theory and the
Schwarzschild metric is the analogous of the constant chromomagnetic field. It
is interesting to note that even in our case, the minimum of the energy
density or the maximum of the induced cosmological constant will be away from
the point $M=0$. It is important to say why one fixes the attention on the
Schwarzschild metric: this is the only metric which is asymptotically flat
depending only by one parameter, the mass $M$, which has zero classical energy
density and therefore it is directly comparable with the Minkowski space. Also
the Anti-de Sitter metric depends only on one parameter: $\Lambda_{AdS}$, but
it is not asymptotically flat and this property makes the comparison between
the two spaces a delicate procedure. The starting point of our analysis will
be the line element $\left(  \ref{dS}\right)  $, which will also be our
cornerstone of the whole paper which is organized as follows. In section
\ref{p1}, we report the results of Refs.\cite{CG,GaMa} and we derive
separately the Hamiltonian constraint for a $f(\mathcal{R})$ theory without
Gravity's Rainbow and the Hamiltonian constraint in presence of Gravity's
Rainbow with the help of the background $\left(  \ref{dS}\right)  $ but
without a $f(\mathcal{R})$ theory term, in section \ref{p2} we compute the ZPE
of quantum fluctuations with an $f(\mathcal{R})$ term around the background
$\left(  \ref{dS}\right)  $ and with the help of an appropriate choice of the
functions $g_{1}\left(  E/E_{P}\right)  $ and $g_{2}\left(  E/E_{P}\right)
$\textbf{,} we will show that the UV divergences of ZPE disappear, in section
\ref{p3} we compute the ZPE of quantum fluctuations with an $f(R)$ term around
the background $\left(  \ref{dS}\right)  $ keeping under control the ZPE UV
divergences. We summarize and conclude in section \ref{p4}. Units in which
$\hbar=c=k=1$ are used throughout the paper.

\section{Setting up the ZPE Computation with the Wheeler-DeWitt Equation}

\label{p1}The semi-classical procedure followed in this section relies heavily
on the formalism outlined in Refs. \cite{CG,Remo1}, where the graviton one
loop contribution to a Schwarzschild background was computed, through a
variational approach with Gaussian trial wave functionals \cite{Remo}. Instead
of using a zeta function regularization to deal with the divergences, we will
use the arbitrariness of the rainbow's functions $g_{1}\left(  E/E_{P}\right)
$ and $g_{2}\left(  E/E_{P}\right)  $ avoiding therefore a renormalization
procedure. Rather than reproduce the formalism, we shall refer the reader to
Refs. \cite{CG,Remo1,GaMa} for details, when necessary. However, for
self-completeness and self-consistency, we present here a brief outline of the
formalism used.

\subsection{The Wheeler-DeWitt Equation in the context of a $f(\mathcal{R})$
gravity theory}

Let us consider now the Lagrangian density describing a generic $f(\mathcal{R}%
)$ theory of gravity, namely
\begin{equation}
\mathcal{L}=\sqrt{-g}\left(  f\left(  \mathcal{R}\right)  -2\Lambda\right)
,\qquad\text{with}\;f^{\prime\prime}\neq0, \label{lag}%
\end{equation}
where $f\left(  \mathcal{R}\right)  $ is an arbitrary smooth function of the
$4D$ scalar curvature and primes denote differentiation with respect to the
scalar curvature. A cosmological term is added also in this case for the sake
of generality. Obviously $f^{\prime\prime}=0$ corresponds to GR. To define the
corresponding Hamiltonian one needs to define a further conjugate
momentum\footnote{See also Ref.\cite{Querella} for technical details.}. To
extract it from the Lagrangian density, we need to write $\mathcal{L}$ in a
form where the Lie derivative $\mathcal{L}_{n}$ is explicit, namely%
\begin{equation}
\mathcal{L}\left[  N,N_{i},g_{ij}\right]  =\sqrt{-\text{\/\thinspace
\thinspace}^{4}\text{\/{}\negthinspace}g}\mathcal{R}=\frac{N{}\,\sqrt{g}%
}{2\kappa}\text{ }\left[  R+K^{2}-3K_{ij}K^{ij}-2N^{-1}g^{ij}N_{|ij}%
-2g^{ij}\mathcal{L}_{n}K_{ij}\right]  .
\end{equation}
Note that in this form boundary terms do not appear. If we define
\begin{equation}
\mathcal{P}^{ij}=N^{-1}\frac{\partial\mathcal{L}}{\partial\left(
\mathcal{L}_{n}K_{ij}\right)  }=-2\sqrt{g}g^{ij}f^{\prime}\left(
\mathcal{R}\right)  \qquad\Longrightarrow\qquad\mathcal{P=}-6\sqrt{g}%
f^{\prime}\left(  \mathcal{R}\right)  , \label{P}%
\end{equation}
then following Ref.\cite{CG}, one finds the generalized Hamiltonian density%
\begin{equation}
\mathcal{H}=\frac{1}{2\kappa}\left[  \frac{\mathcal{P}}{6}\left(  {}%
R-3K_{ij}K^{ij}+K^{2}\right)  +\sqrt{g}U(\mathcal{P})-\frac{1}{3}%
g^{ij}\mathcal{P}_{\mid ij}-2\sqrt{g}K^{ij}K_{ij}\right]  , \label{Hf(R)}%
\end{equation}
where $\Lambda$ is the cosmological constant and
\begin{equation}
U(\mathcal{P})=\mathcal{R}f^{\prime}\left(  \mathcal{R}\right)  -f\left(
\mathcal{R}\right)  \label{V(P)}%
\end{equation}
with the further assumption of having $f^{\prime}\left(  \mathcal{R}\right)
\neq0$. One can use the expression of the canonical momentum $\left(
\ref{mom}\right)  $ to write%
\[
\int_{\Sigma}d^{3}x\left\{  \left[  \left(  2\kappa\right)  G_{ijkl}\pi
^{ij}\pi^{kl}{}-\frac{\sqrt{g}}{2\kappa}{}R\right]  \right.
\]%
\begin{equation}
\left.  +\left(  2\kappa\right)  \left[  G_{ijkl}\pi^{ij}\pi^{kl}+\frac{\pi
}{4}^{2}\right]  \frac{2\left(  f^{\prime}\left(  \mathcal{R}\right)
-1\right)  }{f^{\prime}\left(  \mathcal{R}\right)  }+\frac{\sqrt{g}}{2\kappa
}{}\left(  2\Lambda+\frac{U(\mathcal{P})}{f^{\prime}\left(  \mathcal{R}%
\right)  }\right)  \right\}  =0.
\end{equation}
Defining%
\begin{equation}
h\left(  \mathcal{R}\right)  =1+\frac{2\left[  f^{\prime}\left(
\mathcal{R}\right)  -1\right]  }{f^{\prime}\left(  \mathcal{R}\right)  },
\end{equation}
we can write%
\begin{equation}
\int_{\Sigma}d^{3}x\left\{  \left[  \left(  2\kappa\right)  \left[  h\left(
\mathcal{R}\right)  G_{ijkl}\pi^{ij}\pi^{kl}+\frac{2\left(  f^{\prime}\left(
\mathcal{R}\right)  -1\right)  }{f^{\prime}\left(  \mathcal{R}\right)  }%
\frac{\pi^{2}}{4}\right]  {}-\frac{\sqrt{g}}{2\kappa}{}\left(  R-2\Lambda
-\frac{U(\mathcal{P})}{f^{\prime}\left(  \mathcal{R}\right)  }\right)
\right]  \right\}  =0. \label{GWDW}%
\end{equation}
This is the Hamiltonian constraint, in which we have integrated over the
hypersurface $\Sigma$ and we have used Gauss theorem on the three-divergence.
When $f\left(  \mathcal{R}\right)  =\mathcal{R}$, $U(\mathcal{P})=0$ and%
\begin{equation}
h\left(  \mathcal{R}\right)  =1+\frac{2\left[  f^{\prime}\left(
\mathcal{R}\right)  -1\right]  }{f^{\prime}\left(  \mathcal{R}\right)
}\rightarrow1
\end{equation}
as it should be. The WDW equation $\mathcal{H}\Psi=0$ is the quantum version
of the classical constraint. By repeating the steps which have led to
Eq.$\left(  \ref{VEVO}\right)  $, we obtain\footnote{See Ref.\cite{CG} for
details.}%
\begin{equation}
\frac{\left\langle \Psi\left\vert \int_{\Sigma}d^{3}x\left[  \hat{\Lambda
}_{\Sigma,f\left(  \mathcal{R}\right)  }+2\left(  f^{\prime}\left(
\mathcal{R}\right)  -1\right)  \pi^{2}/\left(  4f^{\prime}\left(
\mathcal{R}\right)  \right)  \right]  \right\vert \Psi\right\rangle
}{V\left\langle \Psi|\Psi\right\rangle }=-\frac{\Lambda^{f\left(
\mathcal{R}\right)  }}{\kappa}, \label{GWDW2}%
\end{equation}
where%
\begin{equation}
\hat{\Lambda}_{\Sigma,f\left(  \mathcal{R}\right)  }=\left(  2\kappa\right)
h\left(  \mathcal{R}\right)  G_{ijkl}\pi^{ij}\pi^{kl}-\frac{\sqrt{g}}{2\kappa
}R
\end{equation}
and where%
\begin{equation}
\Lambda^{f\left(  \mathcal{R}\right)  }=\Lambda+\frac{1}{2V}\int_{\Sigma}%
d^{3}x\sqrt{g}\frac{\mathcal{R}f^{\prime}\left(  \mathcal{R}\right)  -f\left(
\mathcal{R}\right)  }{f^{\prime}\left(  \mathcal{R}\right)  }.
\label{NewLambda}%
\end{equation}
Separating the degrees of freedom to one loop in perturbations, one finds that
the graviton contribution is%
\[
\hat{\Lambda}_{\Sigma,f\left(  \mathcal{R}\right)  }^{\left(  2\right)  ,\bot
}=\frac{1}{4V}\int_{\Sigma}d^{3}x\sqrt{\bar{g}}G^{ijkl}\left[  \left(
2\kappa\right)  h\left(  \mathcal{R}\right)  K^{-1\bot}\left(  x,x\right)
_{ijkl}+\frac{1}{\left(  2\kappa\right)  }\left(  \bigtriangleup_{L\!}%
^{m}K^{\bot}\left(  x,x\right)  \right)  _{ijkl}\right]
\]%
\begin{equation}
=\frac{1}{4}%
{\displaystyle\sum_{\tau}}
{\displaystyle\sum_{i=1}^{2}}
\left[  \left(  2\kappa\right)  h\left(  \mathcal{R}\right)  \lambda
_{i}\left(  \tau\right)  +\frac{E_{i}^{2}\left(  \tau\right)  }{\left(
2\kappa\right)  \lambda_{i}\left(  \tau\right)  }\right]  =-\frac
{\Lambda^{f\left(  \mathcal{R}\right)  }}{\kappa}, \label{LambdaTT}%
\end{equation}
where we have used the following representation for the propagator $K^{\bot
}\left(  x,x\right)  _{iakl}$%
\begin{equation}
K^{\bot}\left(  \overrightarrow{x},\overrightarrow{y}\right)  _{iakl}:=%
{\displaystyle\sum_{\tau}}
\frac{h_{ia}^{\left(  \tau\right)  \bot}\left(  \overrightarrow{x}\right)
h_{kl}^{\left(  \tau\right)  \bot}\left(  \overrightarrow{y}\right)
}{2\lambda\left(  \tau\right)  }%
\end{equation}
and where $h_{ia}^{\left(  \tau\right)  \bot}\left(  \overrightarrow
{x}\right)  $ are the eigenfunctions of $\bigtriangleup_{L\!}^{m}$defined as%
\begin{equation}
\left(  \bigtriangleup_{L\!}^{m}\!{}h^{\bot}\right)  _{ij}=\left(
\bigtriangleup_{L\!}\!{}h^{\bot}\right)  _{ij}-4R{}_{i}^{k}\!{}h_{kj}^{\bot
}+R{}\!{}h_{ij}^{\bot}, \label{M Lichn}%
\end{equation}
which is the modified Lichnerowicz operator where $\bigtriangleup_{L}$is the
Lichnerowicz operator whose expression is%
\begin{equation}
\left(  \bigtriangleup_{L}h\right)  _{ij}=\bigtriangleup h_{ij}-2R_{ikjl}%
h^{kl}+R_{ik}h_{j}^{k}+R_{jk}h_{i}^{k}\qquad\bigtriangleup=-\nabla^{a}%
\nabla_{a}. \label{DeltaL}%
\end{equation}
$\tau$ denotes a complete set of indices and $\lambda\left(  \tau\right)  $
are a set of variational parameters to be determined by the minimization of
Eq.$\left(  \ref{LambdaTT}\right)  $. By minimizing with respect to the
variational function $\lambda_{i}\left(  \tau\right)  $, we obtain the total
one loop energy density for TT tensors%
\begin{equation}
-\frac{\Lambda^{f\left(  \mathcal{R}\right)  }}{\kappa}=\sqrt{h\left(
\mathcal{R}\right)  }\frac{1}{4V}%
{\displaystyle\sum_{\tau}}
\left[  \sqrt{E_{1}^{2}\left(  \tau\right)  }+\sqrt{E_{2}^{2}\left(
\tau\right)  }\right]  . \label{VEVf(R)}%
\end{equation}
The above expression makes sense only for $E_{i}^{2}\left(  \tau\right)  >0$.
Writing explicitly the previous expression we obtain%
\begin{equation}
-\frac{\Lambda}{\kappa}-\frac{1}{2\kappa V}\int_{\Sigma}d^{3}x\sqrt{g}%
\frac{\mathcal{R}f^{\prime}\left(  \mathcal{R}\right)  -f\left(
\mathcal{R}\right)  }{f^{\prime}\left(  \mathcal{R}\right)  }=\sqrt{h\left(
\mathcal{R}\right)  }\frac{1}{4V}%
{\displaystyle\sum_{\tau}}
\left[  \sqrt{E_{1}^{2}\left(  \tau\right)  }+\sqrt{E_{2}^{2}\left(
\tau\right)  }\right]  . \label{modes}%
\end{equation}
To evaluate the right hand side of the previous expression, we need a
regularization prescription. In Ref.\cite{CG} a zeta function regularization
and a renormalization of the induced cosmological constant have been used.
Here we will consider the distortion of the space-time due to Gravity's
Rainbow as a regulator to give meaning to Eq.$\left(  \ref{modes}\right)  $.
To do this we need to show how Gravity's Rainbow enters into the WDW equation.

\subsection{The Wheeler-DeWitt Equation distorted by Gravity's Rainbow}

We refer the reader to Ref. \cite{GaMa} for details, even if a brief outline
will be presented. However, since the Rainbow's functions play a central
r\^{o}le in the whole framework, it is useful to derive how the WDW modifies
when the functions $g_{1}\left(  E/E_{P}\right)  $ and $g_{2}\left(
E/E_{P}\right)  $ distort the background $\left(  \ref{dS}\right)  $. The form
of the background is such that the \textit{shift function}
\begin{equation}
N^{i}=-Nu^{i}=g_{0}^{4i}=0
\end{equation}
vanishes, while $N$ is the previously defined \textit{lapse function}. Thus
the definition of $K_{ij}$ implies
\begin{equation}
K_{ij}=-\frac{\dot{g}_{ij}}{2N}=\frac{g_{1}\left(  E\right)  }{g_{2}%
^{2}\left(  E\right)  }\tilde{K}_{ij}, \label{Kij}%
\end{equation}
where the dot denotes differentiation with respect to the time $t$ and the
tilde indicates the quantity computed in absence of rainbow's functions
$g_{1}\left(  E\right)  $ and $g_{2}\left(  E\right)  $. For simplicity, we
have set $E_{P}=1$ in $g_{1}\left(  E/E_{P}\right)  $ and $g_{2}\left(
E/E_{P}\right)  $ throughout the paragraph. The trace of the extrinsic
curvature, therefore becomes%
\begin{equation}
K=g^{ij}K_{ij}=g_{1}\left(  E\right)  \tilde{K}%
\end{equation}
and the momentum $\pi^{ij}$ conjugate to the three-metric $g_{ij}$ of $\Sigma$
is%
\begin{equation}
\pi^{ij}=\frac{\sqrt{g}}{2\kappa}\left(  Kg^{ij}-K^{ij}\right)  =\frac
{g_{1}\left(  E\right)  }{g_{2}\left(  E\right)  }\tilde{\pi}^{ij}.
\end{equation}
Thus the distorted classical constraint for the $f\left(  \mathcal{R}\right)
=\mathcal{R}$ theory, namely the ordinary GR becomes%
\begin{equation}
\mathcal{H}=\left(  2\kappa\right)  \frac{g_{1}^{2}\left(  E\right)  }%
{g_{2}^{3}\left(  E\right)  }\tilde{G}_{ijkl}\tilde{\pi}^{ij}\tilde{\pi}%
^{kl}\mathcal{-}\frac{\sqrt{\tilde{g}}}{2\kappa g_{2}\left(  E\right)  }%
\!{}\!\left(  \tilde{R}-\frac{2\Lambda_{c}}{g_{2}^{2}\left(  E\right)
}\right)  =0,
\end{equation}
where we have used the following property on $R$%
\begin{equation}
R=g^{ij}R_{ij}=g_{2}^{2}\left(  E\right)  \tilde{R}%
\end{equation}
and where
\begin{equation}
G_{ijkl}=\frac{1}{2\sqrt{g}}\left(  g_{ik}g_{jl}+g_{il}g_{jk}-g_{ij}%
g_{kl}\right)  =\frac{\tilde{G}_{ijkl}}{g_{2}\left(  E\right)  }.
\end{equation}
The symbol \textquotedblleft$\sim$\textquotedblright\ indicates the quantity
computed in absence of rainbow's functions $g_{1}\left(  E\right)  $ and
$g_{2}\left(  E\right)  $. For simplicity, we have set $E_{P}=1$ in
$g_{1}\left(  E/E_{P}\right)  $ and $g_{2}\left(  E/E_{P}\right)  $ throughout
the paragraph. Now we can write the Hamiltonian constraint for the $f\left(
\mathcal{R}\right)  =\mathcal{R}$ theory, namely the ordinary GR%
\begin{equation}
\mathcal{H}=\left(  2\kappa\right)  \frac{g_{1}^{2}\left(  E\right)  }%
{g_{2}^{3}\left(  E\right)  }\tilde{G}_{ijkl}\tilde{\pi}^{ij}\tilde{\pi}%
^{kl}\mathcal{-}\frac{\sqrt{\tilde{g}}}{2\kappa g_{2}\left(  E\right)  }%
\!{}\!\left(  \tilde{R}-\frac{2\Lambda_{c}}{g_{2}^{2}\left(  E\right)
}\right)  =0, \label{Acca}%
\end{equation}
and the corresponding vacuum expectation value $\left(  \ref{VEVO}\right)  $
becomes%
\begin{equation}
\frac{g_{2}^{3}\left(  E\right)  }{\tilde{V}}\frac{\left\langle \Psi\left\vert
\int_{\Sigma}d^{3}x\tilde{\Lambda}_{\Sigma}\right\vert \Psi\right\rangle
}{\left\langle \Psi|\Psi\right\rangle }=-\frac{\Lambda}{\kappa}, \label{WDW1}%
\end{equation}
with%
\begin{equation}
\tilde{\Lambda}_{\Sigma}=\left(  2\kappa\right)  \frac{g_{1}^{2}\left(
E\right)  }{g_{2}^{3}\left(  E\right)  }\tilde{G}_{ijkl}\tilde{\pi}^{ij}%
\tilde{\pi}^{kl}\mathcal{-}\frac{\sqrt{\tilde{g}}\tilde{R}}{\left(
2\kappa\right)  g_{2}\left(  E\right)  }\!{}\!. \label{LambdaR}%
\end{equation}
Extracting the TT tensor contribution from Eq.$\left(  \ref{WDW1}\right)  $,
we find%
\begin{equation}
\hat{\Lambda}_{\Sigma}^{\bot}=\frac{g_{2}^{3}\left(  E\right)  }{4\tilde{V}%
}\int_{\Sigma}d^{3}x\sqrt{\overset{\sim}{\bar{g}}}\tilde{G}^{ijkl}\left[
\left(  2\kappa\right)  \frac{g_{1}^{2}\left(  E\right)  }{g_{2}^{3}\left(
E\right)  }\tilde{K}^{-1\bot}\left(  x,x\right)  _{ijkl}+\frac{1}{\left(
2\kappa\right)  g_{2}\left(  E\right)  }\!{}\left(  \tilde{\bigtriangleup
}_{L\!}^{m}\tilde{K}^{\bot}\left(  x,x\right)  \right)  _{ijkl}\right]  ,
\label{p22}%
\end{equation}
with the prescription that the corresponding eigenvalue equation transforms
into the following way%
\begin{equation}
\left(  \hat{\bigtriangleup}_{L\!}^{m}\!{}h^{\bot}\right)  _{ij}=E^{2}%
h_{ij}^{\bot}\qquad\rightarrow\qquad\left(  \tilde{\bigtriangleup}_{L\!}%
^{m}\!{}\tilde{h}^{\bot}\right)  _{ij}\!{}=\frac{E^{2}}{g_{2}^{2}\left(
E\right)  }\tilde{h}_{ij}^{\bot} \label{EE}%
\end{equation}
in order to reestablish the correct way of transformation of the perturbation.
The propagator $K^{\bot}\left(  x,x\right)  _{iakl}$ will transform as
\begin{equation}
K^{\bot}\left(  \overrightarrow{x},\overrightarrow{y}\right)  _{iakl}%
\rightarrow\frac{1}{g_{2}^{4}\left(  E\right)  }\tilde{K}^{\bot}\left(
\overrightarrow{x},\overrightarrow{y}\right)  _{iakl}. \label{proptt}%
\end{equation}
Thus the total one loop energy density for the graviton for the distorted GR
becomes%
\begin{equation}
\frac{\Lambda}{8\pi G}=-\frac{1}{2\tilde{V}}\sum_{\tau}g_{1}\left(  E\right)
g_{2}\left(  E\right)  \left[  \sqrt{E_{1}^{2}\left(  \tau\right)  }%
+\sqrt{E_{2}^{2}\left(  \tau\right)  }\right]  . \label{VEVR}%
\end{equation}
In the next section we will combine the effects of Gravity's Rainbow and of an
$f\left(  \mathcal{R}\right)  $ theory to see the effects on the induced
cosmological constant.

\section{Gravity's Rainbow and $f\left(  \mathcal{R}\right)  $ gravity at
work}

\label{p2}To combine the effects of Gravity's Rainbow and of an $f\left(
\mathcal{R}\right)  $ theory, we need to know how the scalar curvature
$\mathcal{R}$ transforms when the line element $\left(  \ref{dS}\right)  $ is
considered. We find%
\begin{equation}
\mathcal{R\rightarrow R}_{g_{1}\ g_{2}}=g_{2}^{2}\left(  E\right)  \tilde
{R}+g_{1}^{2}\left(  E\right)  \left(  \tilde{K}_{ij}\tilde{K}^{ij}-\left(
\tilde{K}\right)  ^{2}-2\nabla_{\mu}\left(  \tilde{K}\tilde{u}^{\mu}+\tilde
{a}^{\mu}\right)  \right)  . \label{RR}%
\end{equation}
With the transformation of the scalar curvature $\mathcal{R}$ available, the
new graviton operator becomes%
\[
\hat{\Lambda}_{\Sigma,{}\text{\/ }f\left(  \mathcal{R}_{g_{1}\ g_{2}}\right)
}^{\bot}=\frac{g_{2}^{3}\left(  E\right)  }{4\tilde{V}}\int_{\Sigma}%
d^{3}x\sqrt{\overset{\sim}{\bar{g}}}\left[  \tilde{G}^{ijkl}\left(
2\kappa\right)  \frac{g_{1}^{2}\left(  E\right)  }{g_{2}^{3}\left(  E\right)
}h\left(  \mathcal{R}_{g_{1}\ g_{2}}\right)  \tilde{K}^{-1\bot}\left(
x,x\right)  _{ijkl}\right.
\]%
\begin{equation}
\left.  +\frac{1}{\left(  2\kappa\right)  g_{2}\left(  E\right)  }\!{}\left(
\tilde{\bigtriangleup}_{L\!}^{m}\tilde{K}^{\bot}\left(  x,x\right)  \right)
_{ijkl}\right]  . \label{Lf(R)g1g2}%
\end{equation}
Plugging the form of the propagator into Eq.$\left(  \ref{Lf(R)g1g2}\right)
$, we find%
\begin{equation}
-\frac{\Lambda^{f\left(  \mathcal{R}_{g_{1}\ g_{2}}\right)  }}{\kappa}%
=\frac{1}{2\tilde{V}}%
{\displaystyle\sum_{\tau}}
{\displaystyle\sum_{i=1}^{2}}
g_{2}^{3}\left(  E\right)  \left[  \left(  2\kappa\right)  \frac{g_{1}%
^{2}\left(  E\right)  }{g_{2}^{3}\left(  E\right)  }\lambda_{i}\left(
\tau\right)  +\frac{E_{i}^{2}\left(  \tau\right)  }{\left(  2\kappa\right)
g_{2}\left(  E\right)  \lambda_{i}\left(  \tau\right)  }\right]
\end{equation}
and the minimization with respect to the variational function $\lambda
_{i}\left(  \tau\right)  $ leads to%
\begin{equation}
\frac{\Lambda^{f\left(  \mathcal{R}_{g_{1}\ g_{2}}\right)  }}{\kappa}%
=-\frac{1}{2\tilde{V}}%
{\displaystyle\sum_{\tau}}
\sqrt{h\left(  \mathcal{R}_{g_{1}\ g_{2}}\right)  }g_{1}\left(  E\right)
g_{2}\left(  E\right)  \left[  \sqrt{E_{1}^{2}\left(  \tau\right)  }%
+\sqrt{E_{2}^{2}\left(  \tau\right)  }\right]  , \label{l1loop}%
\end{equation}
where $\Lambda^{f\left(  \mathcal{R}_{g_{1}\ g_{2}}\right)  }$ is expressed by
the Eq.$\left(  \ref{NewLambda}\right)  $ with the obvious replacement
$V\rightarrow\tilde{V}$. The above expression makes sense only for $E_{i}%
^{2}\left(  \tau\right)  >0$. With the help of Regge and Wheeler
representation\cite{Regge Wheeler}, the eigenvalue equation $\left(
\ref{EE}\right)  $ can be reduced to%
\begin{equation}
\left[  -\frac{d^{2}}{dx^{2}}+\frac{l\left(  l+1\right)  }{r^{2}}+m_{i}%
^{2}\left(  r\right)  \right]  f_{i}\left(  x\right)  =\frac{E_{i,l}^{2}%
}{g_{2}^{2}\left(  E\right)  }f_{i}\left(  x\right)  \quad i=1,2\quad,
\label{p34}%
\end{equation}
where we have used reduced fields of the form $f_{i}\left(  x\right)
=F_{i}\left(  x\right)  /r$ and where we have defined two r-dependent
effective masses $m_{1}^{2}\left(  r\right)  $ and $m_{2}^{2}\left(  r\right)
$%
\begin{equation}
\left\{
\begin{array}
[c]{c}%
m_{1}^{2}\left(  r\right)  =\frac{6}{r^{2}}\left(  1-\frac{b\left(  r\right)
}{r}\right)  +\frac{3}{2r^{2}}b^{\prime}\left(  r\right)  -\frac{3}{2r^{3}%
}b\left(  r\right) \\
\\
m_{2}^{2}\left(  r\right)  =\frac{6}{r^{2}}\left(  1-\frac{b\left(  r\right)
}{r}\right)  +\frac{1}{2r^{2}}b^{\prime}\left(  r\right)  +\frac{3}{2r^{3}%
}b\left(  r\right)
\end{array}
\right.  \quad\left(  r\equiv r\left(  x\right)  \right)  . \label{masses}%
\end{equation}
Using the `t Hooft method\cite{tHooft}, we can build the expression of the
modified $\Lambda/\kappa$ which assumes a very complicated
expression\footnote{See the Appendix \ref{app1} for details on the
calculation.}. However, we can obtain enough information if we fix the
attention on some spherically symmetric backgrounds which have the following
property%
\begin{equation}
m_{0}^{2}\left(  r\right)  =m_{2}^{2}\left(  r\right)  =-m_{1}^{2}\left(
r\right)  ,\qquad\forall r\in\left(  r_{t},r_{1}\right)  . \label{cond}%
\end{equation}

For example, the Schwarzschild background represented by the choice $b\left(
r\right)  =r_{t}=2MG$ satisfies the property $\left(  \ref{cond}\right)  $ in
the range $r\in\left[  r_{t},5r_{t}/2\right]  $. Similar backgrounds are the
Schwarzschild-de Sitter and Schwarzschild-Anti de Sitter. On the other hand
other backgrounds, like dS, AdS and Minkowski have the property%
\begin{equation}
m_{0}^{2}\left(  r\right)  =m_{2}^{2}\left(  r\right)  =m_{1}^{2}\left(
r\right)  ,\qquad\forall r\in\left(  r_{t},\infty\right)  . \label{equal}%
\end{equation}
In this paper, we will fix our attention only on metrics which satisfy the
condition $\left(  \ref{cond}\right)  $ and in particular only for the
Schwarzschild background. The spherical symmetry of the metric $\left(
\ref{dS}\right)  $ allows to further reduce the scalar curvature in $4D$,%
\begin{equation}
\mathcal{R}_{g_{1}\ g_{2}}=g_{2}^{2}\left(  E\right)  \tilde{R}=2g_{2}%
^{2}\left(  E\right)  \frac{b^{\prime}\left(  r\right)  }{r^{2}} \label{Rg1g2}%
\end{equation}
where we have used the mixed Ricci tensor $\tilde{R}_{j\text{ }}^{a}$ whose
components are:
\begin{equation}
\tilde{R}_{j\text{ }}^{a}=\left\{  \frac{b^{\prime}\left(  r\right)  }{r^{2}%
}-\frac{b\left(  r\right)  }{r^{3}},\frac{b^{\prime}\left(  r\right)  }%
{2r^{2}}+\frac{b\left(  r\right)  }{2r^{3}},\frac{b^{\prime}\left(  r\right)
}{2r^{2}}+\frac{b\left(  r\right)  }{2r^{3}}\right\}  . \label{Ricci}%
\end{equation}
Thus Eq. $\left(  \ref{NewLf(R)}\right)  $ becomes%
\[
\frac{\Lambda}{\kappa}=-\frac{1}{2\kappa\tilde{V}}\int_{\Sigma}d^{3}x\sqrt
{g}\left[  2g_{2}^{2}\left(  E\right)  \frac{b^{\prime}\left(  r\right)
}{r^{2}}-\frac{f\left(  2g_{2}^{2}\left(  E\right)  \frac{b^{\prime}\left(
r\right)  }{r^{2}}\right)  }{f^{\prime}\left(  2g_{2}^{2}\left(  E\right)
\frac{b^{\prime}\left(  r\right)  }{r^{2}}\right)  }\right]
\]%
\begin{equation}
-\frac{1}{\pi^{2}}\sum_{i=1}^{2}\int_{E^{\ast}}^{+\infty}\sqrt{h\left(
2g_{2}^{2}\left(  E\right)  \frac{b^{\prime}\left(  r\right)  }{r^{2}}\right)
}E_{i}^{2}g_{1}\left(  E\right)  \sqrt{\frac{E_{i}^{2}}{g_{2}^{2}\left(
E\right)  }-m_{i}^{2}\left(  r\right)  }d\left(  \frac{E_{i}}{g_{2}\left(
E\right)  }\right)  \label{Lf(R)g}%
\end{equation}
and for the Schwarzschild background one finds a further simplification%
\begin{equation}
\frac{\Lambda}{\kappa}=\frac{1}{2\kappa}\frac{f\left(  0\right)  }{f^{\prime
}\left(  0\right)  }-\frac{1}{\pi^{2}}\left(  I_{+}+I_{-}\right)  ,
\label{Lf(R)}%
\end{equation}
where%
\begin{equation}
I_{+}=\sqrt{3-\frac{1}{f^{\prime}\left(  0\right)  }}\int_{0}^{+\infty}%
E^{2}g_{1}\left(  E\right)  \sqrt{\frac{E^{2}}{g_{2}^{2}\left(  E\right)
}+m_{0}^{2}\left(  r\right)  }d\left(  \frac{E}{g_{2}\left(  E\right)
}\right)  \label{I+}%
\end{equation}
and%
\begin{equation}
I_{-}=\sqrt{3-\frac{1}{f^{\prime}\left(  0\right)  }}\int_{E^{\ast}}^{+\infty
}E^{2}g_{1}\left(  E\right)  \sqrt{\frac{E^{2}}{g_{2}^{2}\left(  E\right)
}-m_{0}^{2}\left(  r\right)  }d\left(  \frac{E}{g_{2}\left(  E\right)
}\right)  . \label{I-}%
\end{equation}
It is immediate to recognize that if%
\begin{equation}
f^{\prime}\left(  0\right)  =\frac{1}{3}\qquad\Longrightarrow\qquad
I_{+}=I_{-}=0
\end{equation}
and the quantum contribution disappears leaving%
\begin{equation}
\Lambda=\frac{3}{2}f\left(  0\right)  .
\end{equation}
Of course, the disappearance of the quantum contribution to one loop means
that we have to compute higher order contributions to the induced cosmological
constant. On the other hand, when $f^{\prime}\left(  0\right)  \neq1/3$, we
have to evaluate $I_{+}$ and $I_{-}$. To do calculations in practice, we need
to specify the form of $g_{1}\left(  E\right)  $ and $g_{2}\left(  E\right)  $
in such a way $I_{+}$ and $I_{-}$ be finite and condition $\left(
\ref{lim}\right)  $ be satisfied. Since the general case is highly non trivial
as one can deduce from expression $\left(  \ref{Lf(R)g}\right)  $ and it
strongly depends on the form of $f\left(  \mathcal{R}_{g_{1}\ g_{2}}\right)
$, we fix our attention on some examples which can be examined in the context
of the Schwarzschild metric. The examples we are going to discuss are:

\begin{description}
\item[a)]
\begin{equation}
g_{1}\left(  \frac{E}{E_{P}}\right)  =(1+c_{2}\frac{E}{E_{P}})\exp(-c_{1}%
\frac{E^{2}}{E_{P}^{2}})\qquad g_{2}\left(  E/E_{P}\right)  =1, \label{a)}%
\end{equation}

\item[b)]
\begin{equation}
g_{1}\left(  \frac{E}{E_{P}}\right)  =(1+c_{2}\frac{E}{E_{P}})\exp(-c_{1}%
\frac{E^{2}}{E_{P}^{2}})\qquad g_{2}\left(  E/E_{P}\right)  =1+c_{3}\frac
{E}{E_{P}} \label{b)}%
\end{equation}
and

\item[c)] 
\end{description}

\begin{equation}
g_{2}^{-2}(E/E_{P})=g_{1}(E/E_{P}), \label{c)}%
\end{equation}
where for convenience we have reintroduced the dependence on $E_{P}%
$\footnote{Equivalent proposals to models b) and c) from the ultraviolet point
of view are, for example:%
\[
g_{2}\left(  \frac{E}{E_{P}}\right)  =\frac{1}{1+\alpha\frac{E}{E_{P}}%
\tanh\left(  \frac{E}{E_{P}}\right)  }%
\]
and%
\[
g_{2}\left(  \frac{E}{E_{P}}\right)  =\frac{1}{1+\alpha\frac{E}{E_{P}}%
\arctan\left(  \frac{E}{E_{P}}\right)  }.
\]
}. The choice of $g_{1}(E/E_{P})$ proposed in examples $\left(  \ref{a)}%
,\ref{b)}\right)  $ and $\left(  \ref{c)}\right)  $ has been studied
extensively in Refs.\cite{GaMa,GaMa1} and its origin is in a similarity
between the Gravity's Rainbow procedure and the Noncommutative theory analyzed
in Ref.\cite{RGPN}. Anyway, all these cases have to pass the Minkowski limit
test, namely in absence of a curved background one must reproduce a vanishing
cosmological constant. This test furthermore allows to fix the value of
$f\left(  0\right)  $ at least for the Schwarzschild case. Beginning with the
case a), which has been studied in Ref.\cite{GaMa} for $f\left(
\mathcal{R}\right)  =\mathcal{R}$, we find that the computation of the
integrals $\left(  \ref{I+}\right)  $ and $\left(  \ref{I-}\right)  $ leads to
a finite value of the induced cosmological constant. Here, we report the
asymptotic expansion for small and large $x$ of $I_{+}+I_{-}$ with the factor
$\sqrt{3-\frac{1}{f^{\prime}\left(  0\right)  }}$ dropped, where%
\begin{equation}
x=\sqrt{\frac{m_{0}^{2}\left(  r\right)  }{E_{P}^{2}}}=\sqrt{\frac{3MG}%
{r^{3}E_{P}^{2}}} \label{ratio}%
\end{equation}
and where we have used the explicit form of $b\left(  r\right)  =r_{t}=2MG$.
For large $x$, one gets%
\begin{equation}
-\frac{1}{\pi^{2}E_{P}^{4}}\left(  I_{+}+I_{-}\right)  \simeq-{\frac{\left(
2c_{2}c_{1}^{3/2}+\sqrt{\pi}c_{1}^{2}\right)  x}{4\pi^{2}c_{1}^{7/2}}%
-\frac{8c_{2}c_{1}^{5/2}+3\sqrt{\pi}c_{1}^{3}}{16\pi^{2}c_{1}^{11/2}x}%
+\frac{3}{128\pi^{2}}}\,{\frac{16c_{2}c_{1}^{7/2}+5\sqrt{\pi}c_{1}^{4}}%
{c_{1}^{15/2}{x}^{3}}}+O\left(  x^{-4}\right)  , \label{AsL}%
\end{equation}
while for small $x$ we obtain%
\begin{equation}
-\frac{1}{\pi^{2}E_{P}^{4}}\left(  I_{+}+I_{-}\right)  \simeq-{\frac
{4c_{1}^{5/2}+3\sqrt{\pi}c_{2}c_{1}^{2}}{4\pi^{2}c_{1}^{9/2}}}+O\left(
x^{3}\right)  . \label{SmL}%
\end{equation}
It is straightforward to see that if we set%
\begin{equation}
c_{2}=-{\frac{\sqrt{c_{1}\pi}}{2}}, \label{as}%
\end{equation}
the linear divergent term of the asymptotic expansion of expansion $\left(
\ref{AsL}\right)  $ disappears. Plugging the relationship $\left(
\ref{as}\right)  $ into $\left(  \ref{SmL}\right)  $, one finds%
\begin{equation}
\frac{I_{+}+I_{-}}{\pi^{2}E_{P}^{4}}={\frac{3\pi-8}{8\pi^{2}c_{1}^{2}}}.
\end{equation}
Thus the induced cosmological constant $\left(  \ref{Lf(R)}\right)  $ becomes
when $x\rightarrow0$
\begin{equation}
\frac{\Lambda}{\kappa}=\frac{1}{2\kappa}\frac{f\left(  0\right)  }{f^{\prime
}\left(  0\right)  }-E_{P}^{4}\frac{3\pi-8}{8\pi^{2}c_{1}^{2}} \label{Lf(R)a}%
\end{equation}
and using the freedom to fix the value of $f\left(  0\right)  $, we
find\footnote{Note that in Ref.\cite{GaMa}, the value of $c_{1}$ has been
fixed to $1/4$ as suggested by Noncommutative theory\cite{RGPN}.}%
\begin{equation}
f\left(  0\right)  =f^{\prime}\left(  0\right)  2E_{P}^{2}\frac{3\pi-8}{\pi
c_{1}^{2}}\qquad x=0.
\end{equation}
Therefore the appropriate $f\left(  \mathcal{R}\right)  $ model should be
defined with the following property
\begin{equation}
f\left(  0\right)  =\left\{
\begin{array}
[c]{cc}%
f^{\prime}\left(  0\right)  2E_{P}^{2}\left(  3\pi-8\right)  /\left(  \pi
c_{1}^{2}\right)  & x=0\\
& \\
0 & x>0
\end{array}
\right.  . \label{f(0)a}%
\end{equation}
Of course this definition works only for the Schwarzschild background. Note
that when we are on the throat $r=2MG$ $\left(  G=E_{P}^{-2}\right)  $%
\begin{equation}
x=x_{M}=\frac{E_{P}}{2M}\sqrt{\frac{3}{2}}, \label{ratio1}%
\end{equation}
and $x\rightarrow\infty$ when $M\rightarrow0$. However the parametrization
$\left(  \ref{a)}\right)  $ leads to a convergent result when $x\rightarrow
\infty$ as indicated by the expression $\left(  \ref{AsL}\right)  $. Therefore
we conclude that with the choice $\left(  \ref{f(0)a}\right)  $ we can
reproduce the Minkowski limit for every value of $M$. As regards the case b)
we obtain the following result\footnote{See Appendix \ref{app2} for technical
details.}%
\[
\frac{\Lambda}{\kappa}=\frac{1}{2\kappa}\frac{f\left(  0\right)  }{f^{\prime
}\left(  0\right)  }-\frac{E_{P}^{4}}{8\pi^{2}}\sqrt{3-\frac{1}{f^{\prime
}\left(  0\right)  }}\left\{  2\left(  1+x^{2}\right)  ^{\frac{3}{2}}%
-x^{2}\left(  \sqrt{1+x^{2}}+\sqrt{1-x^{2}}\right)  -x^{4}\ln\left(  \frac
{1}{x}+\sqrt{1+\frac{1}{x^{2}}}\right)  \right.
\]%
\begin{equation}
\left.  +2\left(  1-x^{2}\right)  ^{\frac{3}{2}}-x^{4}\ln\left(  \frac{1}%
{x}+\sqrt{1-\frac{1}{x^{2}}}\right)  +\frac{8}{c_{3}^{3}}\left(  \sqrt{\pi
}\left(  1-\,\operatorname{erf}\left(  1/2\right)  \right)  +{\frac{2c_{2}{}%
}{\sqrt{e}}}\right)  \left[  \sqrt{1+c_{3}^{2}x^{2}}+\sqrt{1-c_{3}^{2}x^{2}%
}\right]  \right\}  , \label{Lf(R)b}%
\end{equation}
where we have fixed $c_{1}=1/4$, like in Ref.\cite{GaMa}. Now we have to
verify the Minkowski limit with the computation of%
\begin{equation}
\lim_{x\rightarrow0}\frac{\Lambda}{\kappa}=\frac{1}{2\kappa}\frac{f\left(
0\right)  }{f^{\prime}\left(  0\right)  }-\frac{E_{P}^{4}}{8\pi^{2}}%
\sqrt{3-\frac{1}{f^{\prime}\left(  0\right)  }}\left[  4+\frac{16}{c_{3}^{3}%
}\left(  \sqrt{\pi}\left(  1-\,\operatorname{erf}\left(  1/2\right)  \right)
+{\frac{2c_{2}{}}{\sqrt{e}}}\right)  \right]  =0,
\end{equation}
where $x$ has been defined in $\left(  \ref{ratio}\right)  $. If $f\left(
0\right)  =0$, we have to impose that%
\begin{equation}
c_{2}=-2\sqrt{e}\left(  1+\frac{4}{c_{3}^{3}}\sqrt{\pi}\left(
1-\,\operatorname{erf}\left(  1/2\right)  \right)  \right)
\end{equation}
to obtain the vanishing of the induced cosmological constant $\left(
\ref{Lf(R)b}\right)  $. Otherwise if%
\begin{equation}
f\left(  0\right)  \neq0\qquad\Longrightarrow\qquad f\left(  0\right)
=\frac{2E_{P}^{2}}{\pi}\sqrt{3f^{\prime2}\left(  0\right)  -f^{\prime}\left(
0\right)  }\left[  4+\frac{16}{c_{3}^{3}}\left(  \sqrt{\pi}\left(
1-\,\operatorname{erf}\left(  1/2\right)  \right)  +{\frac{2c_{2}{}}{\sqrt{e}%
}}\right)  \right]  .
\end{equation}
All we have found for the case b) is valid for $r>2MG$. However when we reach
the throat $r_{t}=2MG$, Eq.$\left(  \ref{ratio}\right)  $ has to be
substituted with Eq.$\left(  \ref{ratio1}\right)  $ where for $M\rightarrow0$,
$x\rightarrow\infty$. Therefore we conclude that the case b) cannot be
considered as a good description of the induced cosmological constant for
every value of $M$. As regards the case c), $I_{+}$ and $I_{-}$ reduce to%
\begin{equation}
I_{+}=\sqrt{3-\frac{1}{f^{\prime}\left(  0\right)  }}\int_{0}^{+\infty}\left(
\frac{E}{g_{2}\left(  E/E_{P}\right)  }\right)  ^{2}\sqrt{\frac{E^{2}}%
{g_{2}^{2}\left(  E/E_{P}\right)  }+m_{0}^{2}\left(  r\right)  }d\left(
\frac{E}{g_{2}\left(  E/E_{P}\right)  }\right)
\end{equation}
and%
\begin{equation}
I_{-}=\sqrt{3-\frac{1}{f^{\prime}\left(  0\right)  }}\int_{E^{\ast}}^{+\infty
}\left(  \frac{E}{g_{2}\left(  E/E_{P}\right)  }\right)  ^{2}\sqrt{\frac
{E^{2}}{g_{2}^{2}\left(  E/E_{P}\right)  }-m_{0}^{2}\left(  r\right)
}d\left(  \frac{E}{g_{2}\left(  E/E_{P}\right)  }\right)  .
\end{equation}
With the help of the auxiliary variable%
\begin{equation}
z\left(  E/E_{P}\right)  =\frac{E/E_{P}}{g_{2}\left(  E/E_{P}\right)  },
\end{equation}
we find that Eq.$\left(  \ref{Lf(R)}\right)  $ becomes:%
\begin{equation}
\frac{\Lambda}{8\pi G}=\frac{1}{2\kappa}\frac{f\left(  0\right)  }{f^{\prime
}\left(  0\right)  }-\frac{I\left(  z_{\infty},x\right)  }{8\pi^{2}}E_{P}%
^{4}\sqrt{3-\frac{1}{f^{\prime}\left(  0\right)  }}, \label{Lf(R)g1g24}%
\end{equation}
where $z_{\infty}=\lim\limits_{E\rightarrow\infty}z(E/E_{P})$ and where we
have used Eq.$\left(  \ref{ratio}\right)  $ to obtain%
\[
I\left(  z_{\infty},x\right)  =\left[  z_{\infty}\left(  2z_{\infty}^{2}%
-x^{2}\right)  \sqrt{z_{\infty}^{2}-x^{2}}-x^{4}\ln\left(  \frac{z_{\infty}%
}{x}+\sqrt{\frac{z_{\infty}^{2}}{x^{2}}-1}\right)  \right.
\]%
\begin{equation}
+\left.  z_{\infty}\left(  2z_{\infty}^{2}+x^{2}\right)  \sqrt{z_{\infty}%
^{2}+x^{2}}-x^{4}\ln\left(  \frac{z_{\infty}}{x}+\sqrt{\frac{z_{\infty}^{2}%
}{x^{2}}+1}\right)  \right]  . \label{Intz}%
\end{equation}
For the reasons discussed in case b), we adopt the proposal $\left(
\ref{b)}\right)  $ and the limit%
\begin{equation}
z_{\infty}=\lim\limits_{E\rightarrow\infty}z(\frac{E}{E_{P}})=\lim
\limits_{E\rightarrow\infty}\frac{E/E_{P}}{1+c_{3}E/E_{P}}=\frac{1}{c_{3}}
\label{zinf}%
\end{equation}
becomes a constant. From Eq.$\left(  \ref{Lf(R)g1g24}\right)  $ to verify if
the Minkowski limit is satisfied for the case $r>2MG$, we find%
\begin{equation}
\lim\limits_{M\rightarrow0}\frac{\Lambda}{8\pi G}=\frac{1}{2\kappa}%
\frac{f\left(  0\right)  }{f^{\prime}\left(  0\right)  }-\frac{E_{P}^{4}}%
{8\pi^{2}c_{3}^{4}}\sqrt{3-\frac{1}{f^{\prime}\left(  0\right)  }}I\left(
1,y\right)  . \label{LoGc}%
\end{equation}
where $y=c_{3}x$ and $I\left(  1,y=0\right)  =4$. Thus by defining%
\begin{equation}
f\left(  0\right)  =\frac{f^{\prime}\left(  0\right)  8E_{P}^{2}}{\pi
c_{3}^{4}}\sqrt{3-\frac{1}{f^{\prime}\left(  0\right)  }},
\end{equation}
Eq.$\left(  \ref{LoGc}\right)  $ becomes%
\begin{equation}
\frac{\Lambda}{8\pi G}=\frac{E_{P}^{4}}{8\pi^{2}c_{3}^{4}}\sqrt{3-\frac
{1}{f^{\prime}\left(  0\right)  }}\left[  4-I\left(  1,y\right)  \right]  .
\end{equation}
In this way the Minkowski limit is satisfied. A particular attention has to be
considered for the case $r=2MG$. As shown in Eq.$\left(  \ref{ratio1}\right)
$, $y\rightarrow\infty$ when $M\rightarrow0$. From the limit $\left(
\ref{zinf}\right)  $ and the function $\left(  \ref{Intz}\right)  $, one gets%
\begin{equation}
I\left(  z_{\infty},x\right)  =I\left(  \frac{1}{c_{3}},\frac{E_{P}}{2M}%
\sqrt{\frac{3}{2}}\right)  \label{IzM}%
\end{equation}
which becomes imaginary when $M<c_{3}E_{P}\sqrt{3}/\sqrt{8}$. To avoid this
drawback, we can use the arbitrariness of $c_{3}$ by imposing that $I\left(
a,b\right)  $ be equal to%
\begin{equation}
I\left(  \frac{E_{P}}{2M}\sqrt{\frac{3}{2}},\frac{E_{P}}{2M}\sqrt{\frac{3}{2}%
}\right)  =\frac{9E_{P}^{4}}{64M^{4}}\left[  3\sqrt{2}-\ln\left(  1+\sqrt
{2}\right)  \right]  \qquad\mathrm{when\qquad}M\leq\frac{E_{P}}{2}\sqrt
{\frac{3}{2}}c_{3}. \label{IzMM}%
\end{equation}
In this way, if we impose the Minkowski limit, Eq.$\left(  \ref{LoGc}\right)
$ allows to define%
\begin{equation}
f\left(  0\right)  =f^{\prime}\left(  0\right)  \frac{9E_{P}^{2}}{32\pi
}\left(  \frac{E_{P}}{M}\right)  ^{4}\sqrt{3-\frac{1}{f^{\prime}\left(
0\right)  }}\left(  3\sqrt{2}-\ln\left(  1+\sqrt{2}\right)  \right)  .
\label{f(0)c}%
\end{equation}
Note that in this case we cannot fix $f\left(  0\right)  =0$, because there is
not an appropriate choice of the parameters such that the induced cosmological
constant vanishes. Plugging the value of $f\left(  0\right)  $ of $\left(
\ref{f(0)c}\right)  $ into Eq.$\left(  \ref{LoGc}\right)  $ leads to%
\begin{equation}
\frac{\Lambda}{8\pi G}=\frac{E_{P}^{4}}{8\pi^{2}}\sqrt{3-\frac{1}{f^{\prime
}\left(  0\right)  }}\left[  \frac{9}{64}\left(  \frac{E_{P}}{M}\right)
^{4}\left(  3\sqrt{2}-\ln\left(  1+\sqrt{2}\right)  \right)  -I\left(
\frac{1}{c_{3}},\frac{E_{P}}{2M}\sqrt{\frac{3}{2}}\right)  \right]
\end{equation}
and the Minkowski limit is reached when $M\rightarrow0$.

\section{Gravity's Rainbow and $f\left(  \mathcal{R}\right)  =\mathcal{R}%
+f\left(  R\right)  $ gravity at work}

\label{p3}In the previous section we have considered the effects of Gravity's
Rainbow combined with those of an $f\left(  \mathcal{R}\right)  $ theory on
the computation of the induced cosmological constant. Since we have fixed our
ideas on a Schwarzschild background, the original $f\left(  \mathcal{R}%
\right)  $ theory has considerably reduced to be a function of three scalar
curvature $R$. For this reason, in this section we will consider a theory that
modifies only the three dimensional spatial space. We are therefore led to
consider an arbitrary smooth function of the three dimensional scalar
curvature $f(R)$ combined with an ordinary General Relativity four dimensional
scalar curvature $\mathcal{R}$, namely
\begin{equation}
f\left(  \mathcal{R}\right)  =\mathcal{R}+f\left(  R\right)  . \label{f(R)}%
\end{equation}
In the $\mathcal{ADM}$ formulation, the Lagrangian density becomes%
\begin{equation}
\mathcal{L}=\frac{N}{2\kappa}\sqrt{g}f\left(  \mathcal{R}\right)  =\frac
{N}{2\kappa}\sqrt{g}\left[  R+K_{ij}K^{ij}-\left(  K\right)  ^{2}-2\nabla
_{\mu}\left(  Ku^{\mu}+a^{\mu}\right)  +f\left(  R\right)  \right]  ,
\label{Lagr}%
\end{equation}
where we have used the decomposition $\left(  \ref{R}\right)  $. Differently
from the $f\left(  \mathcal{R}\right)  $ model where%
\begin{equation}
\mathcal{P}^{ij}=-2\sqrt{g}g^{ij}f^{\prime}\left(  \mathcal{R}\right)  ,
\end{equation}
in a $f\left(  R\right)  $ model, $\mathcal{P}^{ij}$ is absent and the
Hamiltonian can be computed in an ordinary way. Therefore the r\^{o}le of
$f\left(  R\right)  $ is to shift the value of $R$. One simply obtains%
\begin{equation}
\mathcal{H}=\left(  2\kappa\right)  G_{ijkl}\pi^{ij}\pi^{kl}\mathcal{-}%
\frac{\sqrt{g}}{2\kappa}\left(  {}f\left(  R\right)  +R-2\Lambda\right)  ,
\end{equation}
where a cosmological term has been introduced for a sake of generality. For a
background of the form $\left(  \ref{dS}\right)  $ in the low energy limit,
the Hamiltonian constraint reduces to%
\begin{equation}
\mathcal{H}={}f\left(  2\frac{b^{\prime}\left(  r\right)  }{r^{2}}\right)
+2\frac{b^{\prime}\left(  r\right)  }{r^{2}}-2\Lambda=0.
\end{equation}
Solutions of the classical constraint depend on a case to case. However for
the simple case of ${}f\left(  R\right)  =const.$, a solution is represented
by a Schwarzschild-de Sitter or Schwarzschild-Anti de Sitter metric depending
on the sign of ${}f\left(  R\right)  -2\Lambda$. In particular, for the
Schwarzschild solution, we find that the classical constraint becomes%
\begin{equation}
{}f\left(  0\right)  =2\Lambda.
\end{equation}
The quantization procedure is obtained with the help of the modified WDW
equation
\begin{equation}
\mathcal{H}\Psi=\left[  \left(  2\kappa\right)  G_{ijkl}\pi^{ij}\pi^{kl}%
-\frac{\sqrt{g}}{2\kappa}\!{}\!\left(  {}f\left(  R\right)  +R-2\Lambda
\right)  \right]  \Psi=0. \label{WDW}%
\end{equation}
Following the procedure which has led to Eq.$\left(  \ref{VEVO}\right)  $, we
can write\cite{CG,Remo1}%
\begin{equation}
\frac{1}{V}\frac{\left\langle \Psi\left\vert \int_{\Sigma}d^{3}x\hat{\Lambda
}_{\Sigma}\right\vert \Psi\right\rangle }{\left\langle \Psi|\Psi\right\rangle
}=-\frac{\Lambda^{f\left(  R\right)  }}{\kappa}, \label{VEV}%
\end{equation}
where we have used Eq.$\left(  \ref{Vol}\right)  $ and Eq.$\left(
\ref{LambdaSigma}\right)  $. In this form, Eq.$\left(  \ref{VEV}\right)  $ can
be used to compute ZPE provided that $\Lambda^{f\left(  R\right)  }/\kappa$ be
considered as an eigenvalue of $\hat{\Lambda}_{\Sigma}$, where in this case%
\begin{equation}
\Lambda^{f\left(  R\right)  }=\Lambda-\frac{1}{2V}\frac{\left\langle
\Psi\left\vert \int_{\Sigma}d^{3}x\sqrt{g}f\left(  R\right)  \right\vert
\Psi\right\rangle }{\left\langle \Psi|\Psi\right\rangle }=\Lambda-\frac{1}%
{2V}\int_{\Sigma}d^{3}x\sqrt{g}f\left(  R\right)  .
\end{equation}
Note that, differently from Eq.$\left(  \ref{NewLambda}\right)  $, the
distorted $\Lambda$ does not depend on derivatives of $f\left(  R\right)  $.
However the evaluation of the l.h.s. of $\left(  \ref{VEV}\right)  $ leads to
the usual one loop divergences. These can be kept under control, with the help
of the Rainbow's functions considered in the previous section. We know that
the line element $\left(  \ref{dS}\right)  $ induces a modification of the
scalar curvature $\mathcal{R}$ leading to $\mathcal{R}_{g_{1}\ g_{2}}$
described by Eq.$\left(  \ref{RR}\right)  $. In this way, the Lagrangian
density $\left(  \ref{Lagr}\right)  $ changes into%
\begin{equation}
\mathcal{L}\rightarrow\frac{\tilde{N}\sqrt{\tilde{g}}}{g_{1}\left(  E\right)
g_{2}^{3}\left(  E\right)  }\left\{  g_{2}^{2}\left(  E\right)  \left[
\tilde{R}+\tilde{K}_{ij}\tilde{K}^{ij}-\left(  \tilde{K}\right)  ^{2}\right]
-2\nabla_{\mu}\left(  g_{1}^{2}\left(  E\right)  \left(  \tilde{K}\tilde
{u}^{\mu}+\tilde{a}^{\mu}\right)  \right)  +f\left(  g_{2}^{2}\left(
E\right)  \tilde{R}\right)  \right\}
\end{equation}
and it is straightforward to see that the modified classical constraint
becomes%
\begin{equation}
\mathcal{H}_{m}=\left(  2\kappa\right)  \frac{g_{1}^{2}\left(  E\right)
}{g_{2}^{3}\left(  E\right)  }\tilde{G}_{ijkl}\tilde{\pi}^{ij}\tilde{\pi}%
^{kl}\mathcal{-}\frac{\sqrt{\tilde{g}}}{2\kappa g_{2}\left(  E\right)  }%
\!{}\!\left(  \tilde{R}+\frac{f\left(  g_{2}^{2}\left(  E\right)  \tilde
{R}\right)  -2\Lambda_{c}}{g_{2}^{2}\left(  E\right)  }\right)  =0,
\label{Hmod}%
\end{equation}
where the \textquotedblleft\ $\sim$\textquotedblright\ symbol means that we
have rescaled every quantity like in section \ref{p2}. The Hamiltonian density
$\left(  \ref{Hmod}\right)  $ on the background $\left(  \ref{dS}\right)  $
simplifies into%
\begin{equation}
\mathcal{H}_{m}={}f\left(  2g_{2}^{2}\left(  E\right)  \frac{b^{\prime}\left(
r\right)  }{r^{2}}\right)  +2\frac{b^{\prime}\left(  r\right)  }{r^{2}}%
-\frac{2\Lambda_{c}}{g_{2}^{2}\left(  E\right)  }=0.
\end{equation}
We can verify that the Schwarzschild solution leads to%
\begin{equation}
g_{2}^{2}\left(  E\right)  {}f\left(  0\right)  =2\Lambda_{c}.
\end{equation}
Following the same steps of the previous section, we can write the graviton
one loop contribution to the induced cosmological constant, whose form is%
\begin{equation}
\frac{\Lambda^{f\left(  R\right)  }}{8\pi G}=-\frac{1}{3\pi^{2}}\sum_{i=1}%
^{2}\int_{E^{\ast}}^{+\infty}E_{i}g_{1}\left(  E\right)  g_{2}\left(
E\right)  \frac{d}{dE_{i}}\sqrt{\left(  \frac{E_{i}^{2}}{g_{2}^{2}\left(
E\right)  }-m_{i}^{2}\left(  r\right)  \right)  ^{3}}dE_{i}. \label{Lambda}%
\end{equation}
Note that with the assumption $\left(  \ref{f(R)}\right)  $, the term
$h\left(  \mathcal{R}_{g_{1}\ g_{2}}\right)  $ is absent and this simplifies
technical calculations and the scalar curvature $R$ can be left unspecified.
Let us see what happens when condition $\left(  \ref{cond}\right)  $ holds. In
this case Eq.$\left(  \ref{Lambda}\right)  $ becomes%
\begin{equation}
\frac{\Lambda^{f\left(  R\right)  }}{8\pi G}=-\frac{1}{3\pi^{2}}\left(
I_{+}+I_{-}\right)  , \label{LoveG}%
\end{equation}
where $I_{+}$ and $I_{-}$ have been defined by Eqs.$\left(  \ref{I+}\right)
$, $\left(  \ref{I-}\right)  $ with the term$\sqrt{3-\frac{1}{f^{\prime
}\left(  0\right)  }}$ dropped. We examine here the same proposals done in
section \ref{p2} for $g_{1}(E/E_{P})$ and $g_{2}(E/E_{P})$. For the case a),
essentially we can repeat what we have done in section \ref{p2}. Following the
steps leading to Eq.$\left(  \ref{Lf(R)a}\right)  $, we find%
\begin{equation}
\lim_{M\rightarrow0}\frac{\Lambda}{8\pi G}=0=\frac{f\left(  0\right)
}{2\kappa}+{\frac{{3\pi-8}}{8\pi^{2}c_{1}^{2}}}E_{P}^{4}%
\end{equation}
or%
\begin{equation}
f\left(  0\right)  =-2{\frac{{3\pi-8}}{\pi c_{1}^{2}}}E_{P}^{2}.
\end{equation}
We are therefore led to define%
\begin{equation}
f\left(  0\right)  =\left\{
\begin{array}
[c]{cc}%
-{\left(  6{\pi-16}\right)  }E_{P}^{2}/\left(  \pi c_{1}^{2}\right)  & x=0\\
& \\
0 & x>0
\end{array}
\right.  .
\end{equation}
This means that the correct Minkowski limit is reached when $M\rightarrow0$
$\forall r\in\left[  r_{t},5r_{t}/2\right]  $. Concerning the case b), if one
repeats the steps leading to Eq.$\left(  \ref{Lf(R)b}\right)  $, also in this
case we conclude that the Minkowski limit cannot be reached for every $r$,
because the value $r=2MG$ with $M\rightarrow0$ cannot vanish. As regards the
case c), we find that Eq.$\left(  \ref{Lambda}\right)  $ becomes:%
\begin{equation}
\frac{\Lambda^{f\left(  R\right)  }}{8\pi G}=-\frac{I\left(  z_{\infty
},x\right)  }{8\pi^{2}}E_{P}^{4}, \label{g1g2inv}%
\end{equation}
where $I\left(  z_{\infty},x\right)  $ has been defined in Eq.$\left(
\ref{Intz}\right)  $. Making the same steps of the case c) one finds%
\begin{equation}
\frac{\Lambda^{f\left(  R\right)  }}{8\pi G}=\frac{\Lambda}{8\pi G}-\frac
{1}{V16\pi G}\int_{\Sigma}d^{3}x\sqrt{g}f\left(  g_{2}^{2}\left(  E\right)
\tilde{R}\right)  =-\frac{E_{P}^{4}}{8\pi^{2}c_{3}^{4}}I\left(  1,y\right)  ,
\label{LfR}%
\end{equation}
where $I\left(  z_{\infty},x\right)  \rightarrow I\left(  1,y\right)  $. By
imposing that the Minkowski limit be satisfied, one finds for $r>2MG$%
\begin{equation}
\frac{\Lambda}{8\pi G}-\frac{f\left(  0\right)  }{16\pi G}=-\frac{E_{P}^{4}%
}{8\pi^{2}c_{3}^{4}}I\left(  1,0\right)  ,
\end{equation}
where for the Schwarzschild solution $\tilde{R}=0$. Then one finds%
\begin{equation}
f\left(  0\right)  =\frac{8E_{P}^{2}}{\pi c_{3}^{4}}, \label{f(0)}%
\end{equation}
where we have used the relationship $I\left(  1,0\right)  =4$. Plugging
$\left(  \ref{f(0)}\right)  $ into Eq.$\left(  \ref{LfR}\right)  $, we find%
\begin{equation}
\frac{\Lambda}{8\pi G}=\frac{E_{P}^{4}}{8\pi^{2}c_{3}^{4}}\left[  4-I\left(
1,y\right)  \right]  ,
\end{equation}
and the Minkowski limit is reached. Concerning the case $r=2MG$, using
Eqs.$\left(  \ref{IzM}\right)  $ and $\left(  \ref{IzMM}\right)  $, we find%
\begin{equation}
\frac{\Lambda}{8\pi G}-\frac{f\left(  0\right)  }{16\pi G}=-\frac{9E_{P}^{8}%
}{512\pi^{2}{M}^{4}}\left[  3\sqrt{2}-\ln\left(  1+\sqrt{2}\right)  \right]  .
\end{equation}
By imposing that%
\begin{equation}
\lim_{M\rightarrow0}\frac{\Lambda}{8\pi G}=0,
\end{equation}
we find%
\begin{equation}
f\left(  0\right)  =\frac{9E_{P}^{6}}{32\pi{M}^{4}}\left[  3\sqrt{2}%
-\ln\left(  1+\sqrt{2}\right)  \right]
\end{equation}
and Eq.$\left(  \ref{LfR}\right)  $ becomes%
\begin{equation}
\frac{\Lambda}{8\pi G}=\frac{E_{P}^{4}}{8\pi^{2}}\left[  \frac{9E_{P}^{4}%
}{64{M}^{4}}\left[  3\sqrt{2}-\ln\left(  1+\sqrt{2}\right)  \right]  -I\left(
\frac{1}{c_{3}},\frac{E_{P}}{2M}\sqrt{\frac{3}{2}}\right)  \right]  .
\end{equation}

\section{Conclusions}

\label{p4}In this paper we have examined the effects of the combination of a
$f\left(  \mathcal{R}\right)  $ theory with Gravity's Rainbow on the
calculation of the induced cosmological constant. With the term
\textquotedblleft\textit{induced}\textquotedblright, we mean that the quantum
fluctuations of the gravitational field generate a ZPE that can be interpreted
as a cosmological constant built exclusively by quantum fluctuations without
the contribution of any matter field. The basic tool is a reinterpretation of
the WDW equation which, in this context, is considered as a vacuum expectation
value defined by Eq.$\left(  \ref{VEVO}\right)  $. This proposal has been
widely explored in a series of papers beginning with Ref.\cite{Remo} where a
zeta function regularization and a following renormalization have been used.
Subsequently a generalization to a $f\left(  \mathcal{R}\right)  $ theory has
been introduced in Ref.\cite{CG}. However, only with the introduction of
Gravity's Rainbow\cite{GaMa,GaMa1}, an alternative method to keep under
control UV divergences can be considered. Therefore, it is quite
straightforward to think that Gravity's Rainbow can be generalized to include
also an $f\left(  \mathcal{R}\right)  $ theory. In this way, one hopes to
capture the UV and the infrared properties in only one model. Since for a
spherically symmetric background, the following identity $\mathcal{R}=R$ can
be used, as shown in Eq.$\left(  \ref{R}\right)  $, we can consider two
different distortions coming from a $f\left(  \mathcal{R}\right)  $ theory:
the full covariant $4D$ scalar curvature $\mathcal{R}$ and the spatial
covariant $3D$ scalar curvature $R$. After having deduced how the respectively
$f\left(  \mathcal{R}\right)  $ models transform under Gravity's Rainbow, we
have obtained the distorted WDW equations. However some of the proposals
presented in Refs.\cite{GaMa,GaMa1} have been discarded because the final
induced cosmological constant had a bad Minkowski limit and a negative value.
Here we have used the arbitrariness of $f\left(  \mathcal{R}\right)  $ and
$f\left(  R\right)  $ to correct the Minkowski limit and the negativity of
$\Lambda/\kappa$. We draw to the reader's attention that for Minkowski limit
we mean the following prescription%
\begin{equation}
\lim_{M\rightarrow0}\frac{\Lambda}{8\pi G}=0 \label{MLim}%
\end{equation}
and not%
\begin{equation}
f\left(  \mathcal{R}\right)  _{|\mathcal{R}=0}=0\qquad\mathrm{or}\qquad
f\left(  R\right)  _{|R=0}=0. \label{Flat}%
\end{equation}
It is important to remark that due to the complexity of transformations of
$f\left(  \mathcal{R}\right)  $ under Gravity's Rainbow, a general analysis is
a very difficult task. This is not the case for a $f\left(  R\right)  $
theory, where in principle the form of the function can be kept quite general,
even if with the choice of a metric of the kind $\left(  \ref{dS}\right)  $,
the complexity of the calculation can be further reduced. A considerable
simplification of the model under examination is obtained for the
Schwarzschild metric where $b\left(  r\right)  =2MG$. With this choice, one
finds $\mathcal{R}=R=0$. Thus the general $f\left(  \mathcal{R}\right)  $
theory reduces to an additional cosmological constant that can be used to
shift the ZPE to the desired Minkowski limit. Indeed for both $f\left(
\mathcal{R}\right)  $ and $f\left(  R\right)  $, we have found that what has
been discarded in Refs.\cite{GaMa,GaMa1}, here can be accepted, provided one
takes into account models admitting a discontinuity on the throat (horizon).
We have to draw the reader's attention on the fact that even if $\mathcal{R}%
=R=0$, this does not mean that the two proposals, namely $f\left(
\mathcal{R}\right)  $ and $f\left(  R\right)  $ merge into one. Indeed the
$f\left(  \mathcal{R}\right)  $ theory generates into the one loop graviton
operator $\left(  \ref{Lf(R)g1g2}\right)  $ a term $h\left(  \mathcal{R}%
_{g_{1}\ g_{2}}\right)  $ containing $f^{\prime}\left(  \mathcal{R}\right)  $
which has its origin into the definition $\left(  \ref{P}\right)  $, while for
a $f\left(  R\right)  $ model, the Lie derivative of the extrinsic curvature
$\mathcal{L}_{n}K_{ij}$ is absent. In summary, for case a) we have%
\begin{equation}
\left\{
\begin{array}
[c]{c}%
f\left(  \mathcal{R}\right)  _{\mathcal{R}=0}=\left\{
\begin{array}
[c]{cc}%
f^{\prime}\left(  0\right)  2E_{P}^{2}\left(  3\pi-8\right)  /\left(  \pi
c_{1}^{2}\right)  & x=0\\
& \\
0 & x>0
\end{array}
\right. \\
\\
f\left(  R\right)  _{|R=0}=\left\{
\begin{array}
[c]{cc}%
-{\left(  6{\pi-16}\right)  }E_{P}^{2}/\left(  \pi c_{1}^{2}\right)  & x=0\\
& \\
0 & x>0
\end{array}
\right.
\end{array}
\right.  ,
\end{equation}
while for case c), one gets%
\begin{equation}
\left\{
\begin{array}
[c]{c}%
f\left(  \mathcal{R}\right)  _{\mathcal{R}=0}=\left\{
\begin{array}
[c]{cc}%
9f^{\prime}\left(  0\right)  E_{P}^{6}\sqrt{3-\frac{1}{f^{\prime}\left(
0\right)  }}\left(  3\sqrt{2}-\ln\left(  1+\sqrt{2}\right)  \right)  /\left(
32\pi M^{4}\right)  & r=2MG\\
& \\
\sqrt{3-\frac{1}{f^{\prime}\left(  0\right)  }}f^{\prime}\left(  0\right)
E_{P}^{2}/\left(  \pi c_{3}^{4}\right)  & r>2MG
\end{array}
\right. \\
\\
f\left(  R\right)  _{|R=0}=\left\{
\begin{array}
[c]{cc}%
9E_{P}^{6}\left[  3\sqrt{2}-\ln\left(  1+\sqrt{2}\right)  \right]  /\left(
32\pi{M}^{4}\right)  & r=2MG\\
& \\
8E_{P}^{2}/\left(  \pi c_{3}^{4}\right)  & r>2MG
\end{array}
\right.
\end{array}
\right.  .
\end{equation}
We recall that the case b) did not produce the correct Minkowski limit for
both $f\left(  \mathcal{R}\right)  $ and $f\left(  R\right)  $ theories and
therefore it has been discarded. At this point one question must be posed:
what is the impact of imposing the Minkowski limit on the behavior of the
\textquotedblleft\textit{cosmological constant}\textquotedblright. First of
all, one has to note that in this approach one finds that what is found is a
\textquotedblleft\textit{dynamical cosmological constant}\textquotedblright%
\ which is variable with the radial coordinate $r$ instead of a time
coordinate $t$. Although this is not a result due to the combined effect of
Gravity's Rainbow and $f\left(  \mathcal{R}\right)  $ or $f\left(  R\right)  $
models, because a \textquotedblleft\textit{dynamical cosmological
constant}\textquotedblright\ was introduced in Ref.\cite{Remo} and
subsequently in Ref.\cite{GaMa}, we have to remark that such a combination
enlarges the family of models that potentially can explain the behavior of the
cosmological constant in the different epochs. For example, the model $\left(
\ref{a)}\right)  $ discussed also in Ref.\cite{GaMa} without the help of a
$f\left(  \mathcal{R}\right)  $ or $f\left(  R\right)  $ modification needed
two different rainbow's functions matching at some space point to have a
correct Minkowski limit for every Schwarzschild mass $M$. Therefore the
combination of Gravity's Rainbow and $f\left(  \mathcal{R}\right)  $ or
$f\left(  R\right)  $ seems to have the right properties to extract the
necessary information about the cosmological constant. We have to stress that
all we need is a finite not vanishing value of $f\left(  \mathcal{R}\right)  $
or $f\left(  R\right)  $ when $\mathcal{R}=R=0$ and a finite value of
$f^{\prime}\left(  \mathcal{R}\right)  $ for $\mathcal{R}=0$. This does not
mean that every proposal can be accepted. For instance models of the form%
\begin{equation}
f\left(  \mathcal{R}\right)  =\mathcal{R}\pm\frac{\mu^{2\left(  n+1\right)  }%
}{\mathcal{R}^{n}}\qquad\mathrm{or}\qquad f\left(  R\right)  =R\pm\frac
{\mu^{2\left(  n+1\right)  }}{R^{n}}\qquad n\geq1 \label{1/R}%
\end{equation}
cannot be taken as a viable examples because of their singularity in the
Schwarzschild background. We arrive to the same conclusion to proposals of the
form\cite{SNSO}%
\begin{equation}
f\left(  \mathcal{R}\right)  =A\ln\left(  \alpha\mathcal{R}\right)
\qquad\mathrm{or}\qquad f\left(  R\right)  =B\ln\left(  \alpha R\right)  ,
\label{ln}%
\end{equation}
where $A$ and $B$ are appropriate constants needed to reestablish the correct
dimensions. At the present stage, we do not know if this is a failure of our
proposal or an indication helping to select the various models. We have also
to remark that the considerations done in $\left(  \ref{1/R}\right)  $ and in
$\left(  \ref{ln}\right)  $ are independent on the Gravity's Rainbow scheme,
at least for the Schwarzschild background. On the other hand proposals like%
\begin{equation}
f\left(  \mathcal{R}\right)  =A\exp\left(  -\alpha\mathcal{R}\right)
\qquad\mathrm{or}\qquad f\left(  R\right)  =B\exp\left(  -\alpha R\right)
\end{equation}
have the correct properties to shift the ZPE solution to the desired Minkowski
value. It is interesting to observe that usually one constrains $f\left(
\mathcal{R}\right)  $ or $f\left(  R\right)  $ to have the flatness property
$\left(  \ref{Flat}\right)  $ and%
\begin{equation}
\lim_{\mathcal{R}\rightarrow\infty}f\left(  \mathcal{R}\right)  =-\Lambda
_{Planck}\qquad\mathrm{or}\qquad\lim_{R\rightarrow\infty}f\left(  R\right)
=-\Lambda_{Planck}%
\end{equation}
to have inflation, where $\Lambda_{Planck}$ is of Planckian size. In our
approach the situation seems to be reversed because the condition $\left(
\ref{MLim}\right)  $ generates a big $f\left(  0\right)  $ in both $4D$ and
$3D$ to compensate the effects of ZPE. At first glance one could conclude that
computing ZPE with the help of a $f\left(  \mathcal{R}\right)  $ combined to
Gravity's Rainbow is not so different to computing ZPE with the help of a
$f\left(  R\right)  $ theory combined with Gravity's Rainbow since the main
difference is in $f^{\prime}\left(  \mathcal{R}\right)  $. However all the
considerations done hitherto are about the Schwarzschild metric, from one side
and from the other side when one introduces also boundary term this difference
is more marked. Indeed the boundary action for a $f\left(  \mathcal{R}\right)
$ model is\cite{EDKH}%
\begin{equation}
\int_{\partial\mathcal{M}}d^{3}x\sqrt{\left\vert h\right\vert }f^{\prime
}\left(  \mathcal{R}\right)  K,
\end{equation}
where $K$ is the trace of the second fundamental form, $\partial\mathcal{M}$
is the boundary of the manifold $\mathcal{M}$ and $h$ is the induced metric on
$\partial\mathcal{M}$. Of course when $f\left(  \mathcal{R}\right)
=\mathcal{R}$, the boundary term reduces to the usual case of General
Relativity which also coincides with the $f\left(  R\right)  $ proposal,
namely%
\begin{equation}
\int_{\partial\mathcal{M}}d^{3}x\sqrt{\left\vert h\right\vert }K.
\end{equation}
This simple but relevant difference opens up a window on the way in which some
problems like black hole pair creation, entropy computation can be computed in
the respective schemes.

\appendix{}

\section{Computing the distorted ZPE in Gravity's Rainbow}

\label{app1}In this Appendix we explicitly derive the expression of the
induced cosmological constant distorted by Gravity's Rainbow and by a
$f\left(  \mathcal{R}\right)  $ theory obtained in Eq.$\left(  \ref{Lf(R)g}%
\right)  $. In order to use the WKB approximation, from Eq.$\left(
\ref{p34}\right)  $ we can extract two r-dependent radial wave numbers%
\begin{equation}
k_{i}^{2}\left(  r,l,\omega_{i,nl}\right)  =\frac{E_{i,nl}^{2}}{g_{2}%
^{2}\left(  E\right)  }-\frac{l\left(  l+1\right)  }{r^{2}}-m_{i}^{2}\left(
r\right)  \quad i=1,2\quad. \label{kTT}%
\end{equation}
The number of modes with frequency less than $E_{i}$, $\left(  i=1,2\right)  $
is given approximately by%
\begin{equation}
\tilde{g}\left(  E_{i}\right)  =\int_{0}^{l_{\max}}\nu_{i}\left(
l,E_{i}\right)  \left(  2l+1\right)  dl, \label{a1}%
\end{equation}
where $\nu_{i}\left(  l,E_{i}\right)  $, $i=1,2$ is the number of nodes in the
mode with $\left(  l,E_{i}\right)  $, such that $\left(  r\equiv r\left(
x\right)  \right)  $
\begin{equation}
\nu_{i}\left(  l,E_{i}\right)  =\frac{1}{\pi}\int_{-\infty}^{+\infty}%
dx\sqrt{k_{i}^{2}\left(  r,l,E_{i}\right)  }. \label{a2}%
\end{equation}
Here it is understood that the integration with respect to $x$ and $l_{\max}$
is taken over those values which satisfy $k_{i}^{2}\left(  r,l,E_{i}\right)
\geq0,$ $i=1,2$. With the help of Eqs.$\left(  \ref{a1},\ref{a2}\right)  $,
Eq.$\left(  \ref{l1loop}\right)  $ becomes%
\begin{equation}
\frac{\Lambda^{f\left(  \mathcal{R}_{g_{1}\ g_{2}}\right)  }}{8\pi G}%
=-\frac{1}{\pi\tilde{V}}\sum_{i=1}^{2}\int_{0}^{+\infty}\sqrt{h\left(
\mathcal{R}_{g_{1}\ g_{2}}\right)  }E_{i}g_{1}\left(  E\right)  g_{2}\left(
E\right)  \frac{d\tilde{g}\left(  E_{i}\right)  }{dE_{i}}dE_{i}, \label{a3}%
\end{equation}
where%
\[
\frac{d\tilde{g}(E_{i})}{dE_{i}}=\int\frac{\partial\nu(l{,}E_{i})}{\partial
E_{i}}(2l+1)dl=\frac{1}{\pi}\int_{-\infty}^{+\infty}dx\int_{0}^{l_{\max}}%
\frac{(2l+1)}{\sqrt{k^{2}(r,l,E)}}\frac{d}{dE_{i}}\left(  \frac{E_{i}^{2}%
}{g_{2}^{2}\left(  E\right)  }-m_{i}^{2}\left(  r\right)  \right)  dl
\]%
\begin{equation}
=\frac{2}{\pi}\int_{-\infty}^{+\infty}dxr^{2}\frac{d}{dE_{i}}\left(
\frac{E_{i}^{2}}{g_{2}^{2}\left(  E\right)  }-m_{i}^{2}\left(  r\right)
\right)  \sqrt{\frac{E_{i}^{2}}{g_{2}^{2}\left(  E\right)  }-m_{i}^{2}\left(
r\right)  }=\frac{4}{3\pi}\int_{-\infty}^{+\infty}dxr^{2}\frac{d}{dE_{i}%
}\left(  \frac{E_{i}^{2}}{g_{2}^{2}\left(  E\right)  }-m_{i}^{2}\left(
r\right)  \right)  ^{\frac{3}{2}}. \label{a4}%
\end{equation}
Plugging expression $\left(  \ref{a4}\right)  $ into Eq.$\left(
\ref{a3}\right)  $, with the help of Eq.$\left(  \ref{Vol}\right)  $ one gets%
\begin{equation}
\int_{-\infty}^{+\infty}dxr^{2}\left[  \frac{\Lambda^{f\left(  \mathcal{R}%
_{g_{1}\ g_{2}}\right)  }}{8\pi G}+\frac{1}{\pi^{2}}\sum_{i=1}^{2}%
\int_{E^{\ast}}^{+\infty}\sqrt{h\left(  \mathcal{R}_{g_{1}\ g_{2}}\right)
}E_{i}^{2}g_{1}\left(  E\right)  \sqrt{\frac{E_{i}^{2}}{g_{2}^{2}\left(
E\right)  }-m_{i}^{2}\left(  r\right)  }d\left(  \frac{E_{i}}{g_{2}\left(
E\right)  }\right)  \right]  =0
\end{equation}
where $E^{\ast}$ is the value which annihilates the argument of the root and
where we have assumed that the effective mass does not depend on the energy
$E$. Using Eq.$\left(  \ref{NewLambda}\right)  $, we find%
\[
\frac{\Lambda}{\kappa}=-\frac{1}{2\kappa\tilde{V}}\int_{\Sigma}d^{3}x\sqrt
{g}\frac{\mathcal{R}f^{\prime}\left(  \mathcal{R}_{g_{1}\ g_{2}}\right)
-f\left(  \mathcal{R}_{g_{1}\ g_{2}}\right)  }{f^{\prime}\left(
\mathcal{R}_{g_{1}\ g_{2}}\right)  }%
\]%
\begin{equation}
-\frac{1}{\pi^{2}}\sum_{i=1}^{2}\int_{E^{\ast}}^{+\infty}\sqrt{h\left(
\mathcal{R}_{g_{1}\ g_{2}}\right)  }E_{i}^{2}g_{1}\left(  E\right)
\sqrt{\frac{E_{i}^{2}}{g_{2}^{2}\left(  E\right)  }-m_{i}^{2}\left(  r\right)
}d\left(  \frac{E_{i}}{g_{2}\left(  E\right)  }\right)  . \label{NewLf(R)}%
\end{equation}

\section{Explicit computation of $I_{+}$ and $I_{-}$ in for the case \ref{b)}}

\label{app2}The integrals $I_{+}$ and $I_{-}$ in Section \ref{p2} can be
separated into two pieces%
\begin{equation}
\left\{
\begin{array}
[c]{c}%
I_{+}=\sqrt{3-\frac{1}{f^{\prime}\left(  0\right)  }}\left(  I_{+,1}%
+I_{+,2}\right) \\
I_{-}=\sqrt{3-\frac{1}{f^{\prime}\left(  0\right)  }}\left(  I_{-,1}%
+I_{-,2}\right)
\end{array}
\right.  ,
\end{equation}
where%
\begin{equation}
\left\{
\begin{array}
[c]{c}%
I_{+,1}=\int_{0}^{E_{P}}E^{2}\sqrt{E^{2}+m_{0}^{2}\left(  r\right)  }dE\\
I_{+,2}=\int_{E_{P}}^{+\infty}E^{2}g_{1}\left(  E\right)  \sqrt{\frac{E^{2}%
}{g_{2}^{2}\left(  E\right)  }+m_{0}^{2}\left(  r\right)  }d\left(  \frac
{E}{g_{2}\left(  E\right)  }\right) \\
\\
I_{-,1}=\int_{\sqrt{m_{0}^{2}\left(  r\right)  }}^{E_{P}}E^{2}\sqrt
{E^{2}-m_{0}^{2}\left(  r\right)  }dE\\
I_{-,2}=\int_{E_{P}}^{+\infty}E^{2}g_{1}\left(  E\right)  \sqrt{\frac{E^{2}%
}{g_{2}^{2}\left(  E\right)  }-m_{0}^{2}\left(  r\right)  }d\left(  \frac
{E}{g_{2}\left(  E\right)  }\right)
\end{array}
\right.  .
\end{equation}
$I_{+,1}$ and $I_{-,1}$ can be computed exactly. Indeed the result of the
integration leads to%
\begin{equation}
\left\{
\begin{array}
[c]{c}%
I_{+,1}=\frac{1}{8}E_{P}^{4}\left[  2\left(  1+x^{2}\right)  ^{\frac{3}{2}%
}-x^{2}\sqrt{1+x^{2}}-x^{4}\left(  \ln\left(  1+\sqrt{x^{2}+1}\right)
-\ln\sqrt{x^{2}}\right)  \right] \\
\\
I_{-,1}=\frac{1}{8}E_{P}^{4}\left[  2\left(  1-x^{2}\right)  ^{\frac{3}{2}%
}-x^{2}\sqrt{1-x^{2}}-x^{4}\left(  \ln\left(  1+\sqrt{1-x^{2}}\right)
-\ln\sqrt{x^{2}}\right)  \right]
\end{array}
\right.  .
\end{equation}
Concerning $I_{+,2}$ and $I_{-,2}$ we get%
\begin{equation}
\left\{
\begin{array}
[c]{c}%
I_{+,2}=\int_{E_{P}}^{+\infty}E^{2}\left(  (1+c_{2}\frac{E}{E_{P}})\exp
(-c_{1}\frac{E^{2}}{E_{P}^{2}})\right)  \sqrt{\left(  \frac{E}{1+c_{3}\frac
{E}{E_{P}}}\right)  ^{2}+m_{0}^{2}\left(  r\right)  }\frac{dE}{\left(
1+c_{3}\frac{E}{E_{P}}\right)  ^{2}}\\
\\
I_{-,2}=\int_{E_{P}}^{+\infty}E^{2}\left(  (1+c_{2}\frac{E}{E_{P}})\exp
(-c_{1}\frac{E^{2}}{E_{P}^{2}})\right)  \sqrt{\left(  \frac{E}{1+c_{3}\frac
{E}{E_{P}}}\right)  ^{2}-m_{0}^{2}\left(  r\right)  }\frac{dE}{\left(
1+c_{3}\frac{E}{E_{P}}\right)  ^{2}}%
\end{array}
\right.  .
\end{equation}
It is immediate to see that for a class of rainbow functions $g_{2}\left(
E/E_{P}\right)  $ increasing faster than $E^{2}$, the integrand in $I_{-,2}$
can become imaginary and therefore leads to an imaginary induced cosmological
constant. This is the main reason to fix the ideas on the choice made in
$\left(  \ref{b)}\right)  $. Note that in the range where $E\in\left[
E_{P},+\infty\right)  $, one can write $g_{2}\left(  E/E_{P}\right)  \sim
E_{P}/\left(  c_{3}E\right)  $. Then using the definition $\left(
\ref{ratio}\right)  $, $I_{+,2}$ and $I_{-,2}$ can be easily calculated to
give%
\begin{equation}
\left\{
\begin{array}
[c]{c}%
I_{+,2}=\frac{E_{P}^{4}}{2\sqrt{c_{1}}c_{3}^{3}}\sqrt{1+c_{3}^{2}x^{2}}\left[
\sqrt{\pi}\left(  1-\,\operatorname{erf}\left(  \sqrt{c_{1}}\right)  \right)
+c_{2}{\frac{e^{-c_{1}}{}}{\sqrt{c_{1}}}}\right] \\
\\
I_{-,2}=\frac{E_{P}^{4}}{2\sqrt{c_{1}}c_{3}^{3}}\sqrt{1-c_{3}^{2}x^{2}}\left[
\sqrt{\pi}\left(  1-\,\operatorname{erf}\left(  \sqrt{c_{1}}\right)  \right)
+c_{2}{\frac{e^{-c_{1}}{}}{\sqrt{c_{1}}}}\right]
\end{array}
\right.  ,
\end{equation}
where $\operatorname{erf}\left(  x\right)  $ is the error function and where
we have used the following relationship%
\begin{equation}
\int_{E_{P}}^{+\infty}\left[  (1+c_{2}\frac{E}{E_{P}})\exp(-c_{1}\frac{E^{2}%
}{E_{P}^{2}})\right]  dE={{\frac{E_{P}\sqrt{\pi}}{2\sqrt{c_{1}}}}}\left(
1-\,\operatorname{erf}\left(  \sqrt{c_{1}}\right)  \right)  +c_{2}%
{\frac{e^{-c_{1}}{}}{2c_{1}}.}%
\end{equation}

\end{document}